\documentclass{emulateapj}

\newcommand{\Msol}{$M_{\odot}$}
\newcommand{\be}{\begin{equation}}
\newcommand{\ee}{\end{equation}}
\newcommand{\kms}{\mbox{km s$^{-1}$}}
\newcommand{\hh}{h$^{-1}$}

\begin{document}


\title{The nonlinear biasing of the 10k zCOSMOS galaxies up to $z \sim 1$\footnotemark[1]}


\author{
K.~Kova\v{c}\altaffilmark{2},
C.~ Porciani\altaffilmark{2,3},
S. J.~Lilly\altaffilmark{2},
C.~ Marinoni\altaffilmark{4},
L.~ Guzzo\altaffilmark{5},
O.~ Cucciati\altaffilmark{6},
G.~ Zamorani\altaffilmark{7},
A.~ Iovino\altaffilmark{5},
P.~ Oesch\altaffilmark{2},
M.~Bolzonella\altaffilmark{7},
Y.~ Peng\altaffilmark{2},
B.~ Meneux\altaffilmark{8,9},
E.~ Zucca\altaffilmark{7},
S.~Bardelli\altaffilmark{7},
C. M.~Carollo\altaffilmark{2},
T.~Contini\altaffilmark{10},
J.-P.~ Kneib\altaffilmark{6},
O.~ Le F\`{e}vre\altaffilmark{6},
V.~Mainieri\altaffilmark{11},
A.~ Renzini\altaffilmark{12},
M.~ Scodeggio\altaffilmark{13},
A.~Bongiorno\altaffilmark{8},
K.~Caputi\altaffilmark{2},
G.~ Coppa\altaffilmark{7},
S.~ de la Torre\altaffilmark{6,5,13},
L.~ de Ravel\altaffilmark{6},
A. ~Finoguenov\altaffilmark{8},
P.~ Franzetti\altaffilmark{13},
B.~ Garilli\altaffilmark{13},
P.~ Kampczyk\altaffilmark{2},
C. ~Knobel\altaffilmark{2},
F.~Lamareille\altaffilmark{10},
J.-F.~ Le Borgne\altaffilmark{10},
V.~ Le Brun\altaffilmark{6},
C.~ Maier\altaffilmark{2},
M.~ Mignoli\altaffilmark{7},
R.~ Pello\altaffilmark{10},
E.~ Perez Montero\altaffilmark{10},
L.~ Pozzetti\altaffilmark{7},
E.~ Ricciardelli\altaffilmark{14},
J.~D.~Silverman\altaffilmark{2},
M.~ Tanaka\altaffilmark{11},
L.~A.~M.~ Tasca\altaffilmark{6,13},
L.~ Tresse\altaffilmark{6},
D.~ Vergani\altaffilmark{7},
U.~ Abbas\altaffilmark{6,15},
D.~ Bottini\altaffilmark{13},
A.~ Cappi\altaffilmark{7},
P.~ Cassata\altaffilmark{6,16},
A.~ Cimatti\altaffilmark{17},
M.~ Fumana\altaffilmark{13},
A.~M.~ Koekemoer\altaffilmark{18},
A.~ Leauthaud\altaffilmark{19},
D.~ Maccagni\altaffilmark{13},
H.~ J. McCracken\altaffilmark{20},
P.~ Memeo\altaffilmark{13},
R.~ Scaramella\altaffilmark{21},
N.~Z.~ Scoville\altaffilmark{22}
}

\altaffiltext{2}{Institute of Astronomy, ETH Zurich, 8093 Zurich, Switzerland; kovac@phys.ethz.ch}
\altaffiltext{3}{Argelander Institut f\"ur Astronomie, Auf dem H\"ugel 71, D-53121 Bonn,Germany}
\altaffiltext{4}{Centre de Physique Theorique, Marseille, Marseille,  
France}
\altaffiltext{5}{INAF Osservatorio Astronomico di Brera, Milan, Italy}
\altaffiltext{6}{Laboratoire d'Astrophysique de Marseille, Marseille,  
France}
\altaffiltext{7}{INAF Osservatorio Astronomico di Bologna, via  
Ranzani 1, I-40127, Bologna, Italy}
\altaffiltext{8}{Max-Planck-Institut f\"ur extraterrestrische Physik,  
D-84571 Garching, Germany}
\altaffiltext{9}{Universitats-Sternwarte, Scheinerstrasse 1, D-81679  
Muenchen, Germany}
\altaffiltext{10}{ Laboratoire d'Astrophysique de Toulouse-Tarbes, Universite de Toulouse, CNRS, 14 avenue Edouard Belin, F-31400 Toulouse, France}
\altaffiltext{11}{European Southern Observatory, Karl-Schwarzschild- 
Strasse 2, Garching, D-85748, Germany}
\altaffiltext{12}{INAF - Osservatorio Astronomico di Padova, Padova, Italy}
\altaffiltext{13}{INAF - IASF Milano, Milan, Italy}
\altaffiltext{14}{Dipartimento di Astronomia, Universita di Padova,  
Padova, Italy}
\altaffiltext{15}{INAF Osservatorio Astronomico di Torino, Strada Osservatorio 20, I-10025 Pino Torinese, Torino, Italy}
\altaffiltext{16}{Dept. of Astronomy, University of Massachusetts at Amherst}
\altaffiltext{17}{Dipartimento di Astronomia, Universit\'a di Bologna,  
via Ranzani 1, I-40127, Bologna, Italy}
\altaffiltext{18}{Space Telescope Science Institute, 3700 San Martin Drive, Baltimore, MD 21218}
\altaffiltext{19}{Physics Division, MS 50 R5004, Lawrence Berkeley National Laboratory, 1 Cyclotron Rd., Berkeley, CA 94720, USA}
\altaffiltext{20}{Institut d'Astrophysique de Paris, UMR 7095 CNRS,  
Universit\'e Pierre et Marie Curie, 98 bis Boulevard Arago, F-75014  
Paris, France.}
\altaffiltext{21}{INAF, Osservatorio di Roma, Monteporzio Catone  
(RM), Italy}
\altaffiltext{22}{California Institute of Technology, MS 105-24,
Pasadena, CA 91125, USA}

\footnotetext[1]{Based on observations
   obtained at the European Southern Observatory (ESO) Very Large
   Telescope (VLT), Paranal, Chile, as part of the Large Program
   175.A-0839 (the zCOSMOS Spectroscopic Redshift Survey)}






\begin{abstract}  We use  the overdensity  field reconstructed  in the
volume  of the  COSMOS area  to study  the nonlinear  biasing  of the
zCOSMOS galaxies. The galaxy  overdensity field is reconstructed using
the current sample of $\sim$8500 accurate zCOSMOS redshifts at $I_{AB}
<  22.5$ out  to z$\sim$1  on scales  $R$  from 8  to 12  \hh Mpc.  By
comparing  the  probability  distribution  function  (PDF)  of  galaxy
density contrast $\delta_g$ to  the lognormal approximation of the PDF
of  the mass  density contrast  $\delta$, we  obtain the  mean biasing
function $b(\delta,  z,R)$ between  the galaxy and  matter overdensity
field  and  its  second  moments  $\hat{b}$  and
$\tilde{b}$ up  to  $z  \sim  1$. Over  the redshift  interval $0.4<z<1$ the conditional
mean function $\langle \delta_g|\delta \rangle  = b(\delta,  z,R) \delta$  is of the following characteristic shape. The function  vanishes in the most underdense  regions and then
sharply  rises in  a  nonlinear  way towards  the  mean densities. $\langle \delta_g|\delta \rangle$ is almost a linear tracer of the matter in the overdense regions,
up to  the most overdense regions  in which it is  nonlinear again and
the local effective slope of $\langle \delta_g|\delta \rangle$ vs. $\delta$  is smaller than unity. The $\langle \delta_g|\delta \rangle$ function is evolving  only  slightly over  the
redshift  interval $0.4<z<1$. The  linear biasing  parameter increases
from $\hat{b}=1.24  \pm 0.11$ at $z=0.4$ to  $\hat{b}=1.64 \pm 0.15$ at  $z=1$  for  the  $M_B<-20-z$  sample  of
galaxies. $\hat{b}$ does  not show  any  dependence on  the smoothing  scale from 8 to 12 \hh Mpc,  but increases  with  luminosity.  The
measured nonlinearity  parameter $\tilde{b}/ \hat{b}$ is  of the order
of a  few percent (but it  can be consistent  with 0) and it  does not
change  with  redshift, the smoothing  scale or  the
luminosity. By matching the linear  bias of galaxies to the halo bias,
we infer  that the $M_B<-20-z$  galaxies reside in dark  matter haloes
with  a  characteristic mass  of  about  $3-6  \times 10^{12}$  \Msol,
depending on the halo bias fit.

\end{abstract}

\section{Introduction}

The large scale structure in the Universe is believed to have formed via
gravitational instability of small, primordial density fluctuations.
Virialised dark matter haloes are produced by the collapse of some overdense 
regions
followed by hierarchical merging. 
Galaxies are formed within the dark matter haloes through multiplex
processes including gas cooling, star formation and feedback which are
difficult to model accurately \citep[e.g.][]{White&Rees.1978}.
It is thus expected that the relation between galaxies and the underlying
matter distribution will be also complex. In particular, 
the efficiency of galaxy formation and the rate of galaxy evolution might vary
from place to place depending on the matter density field. 
Therefore the actual galaxy distributions is not expected to be a rightful
tracer of the underlying mass. 
This phenomenon is often referred to as ``galaxy biasing''. 

In order to extract cosmological information from galaxy surveys
it is important to model and parameterise galaxy biasing. This often
requires using a statistical approach. Assuming that galaxies preferentially form within the peaks of the
primordial density distribution,
\citet{Kaiser.1984} showed that 
the two-point correlation function of the galaxy distribution,
 $\xi_{gg}$, should be amplified with respect to the mass autocorrelation function,
$\xi_{mm}$, according to the relation:
\be \xi_{gg}(r) = b_{\xi}^2 \xi_{mm}(r) 
\label{eq_biasksi}
\ee
where the ``biasing parameter'' $b_{\xi}$ is independent of the spatial separation $r$. A similar relation is obtained by relating
the density contrast of the matter $\delta$ and of the galaxies $\delta_g$ at some position $\bf{r}$ through the deterministic and linear relation 
\be  \delta_g ({\bf r}) = b\, \delta ({\bf r}) ~. 
\label{eq_biasdetlin}
\ee
This is the simplest model for galaxy biasing and is still commonly used.
While equation~\ref{eq_biasksi} follows from equation~\ref{eq_biasdetlin}, the
opposite does not hold. 
An obvious deficiency in the definition of $\delta_g$ above is that it will break down in
the most underdense regions $\delta\ll0$ if $b>1$, as values of
$\delta_g<-1$ are not possible. 
This implies that the galaxy bias $b$ must be a nonlinear function of
$\delta$ and, in general, it can also vary with redshift $z$, 
galaxy type and the smoothing scale $R$
used to define the density contrast:
\be b=b(\delta, z, R).  \ee
\citet{Fry&Gaztanaga.1993} proposed to parameterise this function in terms of 
coefficients of the Taylor expansion
\be
\delta_g=b_0+b_1\,\delta+\frac{b_2}{2}\,\delta^2+\dots  \;,
\label{eq_biasfg}
\ee
which are not fully independent as the conditions $\langle \delta_g \rangle=0$
and $\delta_g(\delta=-1)=-1$ must hold.

Galaxy biasing  is  also expected  to  have a stochastic element:  for  any  given  value
of $\delta$ there will be a whole distribution of values for $\delta_g$. 
The stochasticity originates from a number of different sources. First, 
the dynamics of large scale flows 
depends on extra variables beyond the value of the local density
contrast $\delta$ (e.g. on the tidal tensor) 
and makes the bias relation nonlinear, non-local and
stochastic \citep{Catelan.etal.1998}. 
Second, the efficiency of galaxy formation depends
on details of the gas physics. Third, galaxies are discrete objects and any
attempt to reconstruct $\delta_g$ will be effected by shot noise.

\citet{Dekel&Lahav.1999} have  proposed a
formalism  which  separately accounts for the nonlinearity  and  
stochasticity  of  the biasing process. 
Galaxy biasing is described in terms of the conditional probability function
$P(\delta_g|\delta)$ and its moments.
A key quantity here is the mean biasing function $b(\delta)$ defined by
the conditional mean:
\be  
b(\delta) \delta = \langle \delta_g | \delta \rangle = \int d \delta_g P(\delta_g|\delta) \delta_g .   
\ee
%
The mean biasing function $b(\delta$) and its nonlinearity can be 
characterised by its second non-trivial moments:
\be  \hat{b} \equiv \frac {\langle b(\delta)\delta^2 \rangle}{\sigma^2} \ee
and
\be \tilde{b}^2 \equiv \frac {\langle b^2(\delta)\delta^2 \rangle}{\sigma^2} , \ee 
with  $\sigma^2$ the variance of the mass density contrast distribution. 
The parameter $\hat{b}$ measures the slope of the linear regression of 
$\delta_g$ against $\delta$. In the case of linear biasing (see equation~\ref{eq_biasdetlin}), 
both $\hat{b}$ and $\tilde{b}$ 
reduce to the constant bias. 
The ratio $\tilde{b}/\hat{b}$ is thus a measure of the nonlinearity in the 
biasing relation. 
Moreover,
the local variance of $\delta_g$ at fixed $\delta$, $\sigma^2_g(\delta)$ 
can be used to quantify
the degree of stochasticity of the biasing relation.

Based on the Press-Schechter formalism and its extensions
\citep{Bond.etal.1991}, \citet{Mo&White.1996} developed an analytical model 
for the mean biasing relation of the dark matter haloes. This assumes 
that large scale motions follow the spherical collapse 
approximation. The general case is discussed
by \citet{Catelan.etal.1998}.
Related work has been presented in \citet{Mo.etal.1997} 
and Porciani et al. (1998, see also Scannapieco \& Barkana 2002)
\nocite{Porciani.etal.1998, Scannapieco&Barkana.2002}
where two-point and higher-order statistics are considered. Following the analytical approach by \citet{Mo&White.1996},
a number of studies based on N-body simulations were
carried out to study the halo bias \citep[e.g.][]{Jing.1998, Porciani.etal.1999,  Sheth&Lemson.1999, Sheth&Tormen.1999, Jing.1999,
  Kravtsov&Klypin.1999, Sheth.etal.2001, Seljak&Warren.2004, Tinker.etal.2005,
  Pillepich.etal.2008}, leading to a new set of fitting formulae for the
mean biasing relation, 
and a better understanding of the origin of halo biasing.
Independently of the exact halo
definition, assumed cosmology, simulation box size and resolution, 
there is a consensus that in a cold-dark matter scenario: i) at a given epoch,
more massive haloes are more biased tracers of the underlying 
matter than lower mass haloes; ii) for halos of fixed mass, the amount of 
biasing increases with redshift.

However, it is still a huge step 
from a successful description of ``halo biasing'' to that of ``galaxy biasing''
as the latter requires
incorporating a recipe for galaxy formation (and evolution) within the 
current cosmological framework.
Galaxy biasing has been studied through hydrodynamic simulations
\citep[e.g.][]{Blanton.etal.1999, Blanton.etal.2000, Cen&Ostriker.2000,
Yoshikawa.etal.2001} and
semi-analytical modelling combined with N-body simulations
\citep[e.g.][hereafter SBD]{Kauffmann.etal.1997, 
Benson.etal.2000, Somerville.etal.2001,
  Sigad.etal.2000}. 
Despite the difference in the treatment of the various gas-related processes, 
all these studies reach the following consistent conclusions: galaxy biasing is expected to be nonlinear, 
to depend on the properties of the considered galaxy sample 
(and of their host dark matter haloes), and to be a function of cosmic time.

There is now lot of observational evidence for galaxy biasing, 
and its dependence on galaxy type or redshift. 
The Dressler's morphology-density relation \citep{Dressler.1980} 
-- the observational evidence that early type galaxies are more abundant 
in dense regions than spiral galaxies -- is a textbook example for this. 
Building on this, one can summarise decades of observations 
in the local universe 
with a statement that bulge-dominated, red galaxies with mainly 
old stellar populations preferentially live in dense regions, 
and they are  more strongly clustered than the disk-dominated, blue, 
young galaxies \citep[e.g.][]{Norberg.etal.2001, Norberg.etal.2002, Zehavi.etal.2005}. 
Consistently,
deriving the bias parameter from clustering studies 
(see e.g. Equation~\ref{eq_biasksi}) suggests that early-type galaxies 
have a higher bias than late-type galaxies at all luminosities 
\citep{Norberg.etal.2002}. 
On the other hand,
HI-selected galaxies have some of the lowest bias values of all known 
objects \citep{Basilakos.etal.2007}. 
At higher redshifts, galaxies are more biased tracers of 
matter, 
with linear bias parameters of $b \sim 1.48$ at $0.7 < z < 1.3$ 
for $M_B<-20$ galaxies (\citealt{Coil.etal.2006}, see also \citealt{Pollo.etal.2006} and \citealt{Meneux.etal.2009}), 
up to the highest biased samples of extremely red objects (EROs), Lyman Break galaxies (LBGs) and 
Lyman-$\alpha$ emitters. 
For example, the bias parameter
of EROs at $z=1.2$ is $b=2.7$ \citep{Moustakas&Somerville.2002}, LBGs  at $z \sim 3.8$ and $z \sim 4.9$ is $b \sim  2.5$ and $b \sim 4$, respectively \citep{Lee.etal.2006} and the bias parameter of 
Lyman-$\alpha$ emitters at $z \sim 4.5$ 
is $b \sim 3.7$ \citep{Kovac.etal.2007}. 
However, inferring the exact redshift evolution of the biasing process 
from the observational 
data is not straightforward, because galaxy surveys typically sample  
different populations of galaxies at different redshifts 
\citep[e.g.][]{Kovac.etal.2007}.  

A simple way to model the biasing of galaxies which has received 
a lot of attention recently, is through the halo occupation distribution (HOD) 
formalism \citep[e.g. see review by][and references therein]{Cooray&Sheth.2002}. 
This method splits the bias problem into two steps:
i) N-body simulations are used to characterise the spatial distribution
and the clustering properties of virialised dark matter halos as a function of 
their mass (and/or some other properties) for a given cosmology; 
ii) the galaxy distribution
is described in terms of the probability distribution that a halo of mass $M$ 
hosts $N$ galaxies of a specified type (the HOD).
The first N moments of the HOD can be measured by fitting observed
N-point statistics.  
Using the two-point correlation function,
this approach has been widely employed to estimate the mass of the host
halos of galaxies and quasars at low and high redshift 
\citep[e.g.][]{Magliocchetti&Porciani.2003, Porciani.etal.2004, Abazajian.etal.2005, Zehavi.etal.2005, Phleps.etal.2006, Zheng.etal.2007} and also to 
derive the mass-to-light ratio of virialised cosmic structures
\citep[e.g.][]{Yang.etal.2005, Tinker.etal.2005, vandenBosch.etal.2007}.
 
It is however interesting to go beyond the measurement of a (possibly
scale-dependent) bias parameter from two-point statistics.
\citet{Tegmark&Bromley.1999} presented evidence that the present-day 
galaxy biasing is nonlinear and stochastic, employing the galaxy clustering 
in Las Campanas Redshift Survey. 
Using the 2 degree Field Galaxy Redshift Survey (2dFGRS) data, 
some contradictory results on the nonlinear nature of galaxy bias 
have been obtained. 
While \citet{Verde.etal.2002} found no significant evidence for 
nonlinearity from the bispectrum analysis, 
\citet{Gaztanaga.etal.2005} detect non-vanishing 
quadratic corrections in three-point correlation function.
Moreover,
\citet{Wild.etal.2005} and  \citet{Conway.etal.2005} 
using the count-in-cells analysis exclude the deterministic linear bias model 
in both the flux limited and volume limited (luminosity complete) 
samples of galaxies and find  evidence for stochasticity. 
Measurements of the three-point correlation function and counts-in-cells
statistics for galaxy samples from the Sloan Digital Sky Survey (SDSS)
suggest that galaxy biasing is nonlinear and fairly complex 
\citep{Kayo.etal.2004, Nishimichi.etal.2007, Swanson.etal.2008}.

The mean nonlinear biasing function of a sample of galaxies can be constrained by
combining counts-in-cells measurements with models for 
the probability distribution function (PDF) of mass density fluctuations
(SBD, \citealt{Szapudi&Pan.2004}). \citet{Marinoni.etal.2005} applied this technique to the first-epoch
VIMOS  VLT deep survey \citep[VVDS][]{LeFevre.etal.2005},  
over  the  redshift range 0.4-1.5  
on a characteristic scale  $R$ from 5 to 10 \hh Mpc. 
They conclude that galaxy bias increases with redshift and 
is nonlinear in all redshift bins probed. 
In addition, brighter galaxies are more strongly biased than
less luminous ones, as well as redder galaxies
are more biased than blue ones independently of redshift.
However, the area covered by the VVDS used for this study was rather small ($0.4\times0.4$ deg$^2$)
and the results are likely affected by cosmic variance.

It is therefore very important to cross-check their robustness against 
richer datasets. With this spirit,
in this paper we perform a similar analysis on a larger sample of high redshift
galaxies. We make use of  
the first $\sim$ 10000 spectra of the zCOSMOS redshift survey 
\citep{Lilly.etal.2009} to measure galaxy densities within top-hat spheres
of radius 8-12 \hh Mpc 
in the redshift range 0.4-1 \citep[see][]{Kovac.etal.2009}.
Assuming concordance cosmology, we then use 
the PDF of the galaxy density contrast to estimate 
the mean bias function $b=b(\delta, z, R)$ between zCOSMOS galaxies and 
matter overdensities. Within the limits of the zCOSMOS survey, we explore 
the shape and the luminosity, scale and redshift dependence of the conditional mean function $\langle \delta_g|\delta \rangle$. A careful analysis of random and systematic errors based on mock galaxy
catalogues is also presented. Finally, the characteristic mass
of the halos hosting zCOSMOS galaxies is inferred from the value of the linear
bias parameter. The structure of this paper is as follows. We present the data used for the analysis and describe the method to reconstruct the density field in Section~\ref{sec_data}. In Section~\ref{sub_method} we provide the details of the method which we adopt to derive the biasing function. We present the results of the tests on the mock catalogues in Section~\ref{sec_tests}. Finally, we present the biasing analysis on the real data set in Section~\ref{sec_analysis}, comparison to the literature results in Section~\ref{sec_comparison}  and conclusions in Section~\ref{sec_concl}.

The assumed cosmology is specified by $\sigma_8=0.8$, $\Omega_{m,0}=0.25$, $\Omega_{\Lambda,0}=0.75$ and $H_0=70$ \kms Mpc$^{-1}$. We present the obtained results scaled to the units of dimensionless $h$ parameter, $H_0=100 h$ km s$^{-1}$ Mpc$^{-1}$, except of magnitudes, which are calculated with $h=0.7$.

\section{The zCOSMOS survey}
\label{sec_data}

\subsection{The 10k sample}

zCOSMOS \citep{Lilly.etal.2007, Lilly.etal.2009} is a redshift survey undertaken in the field of the multiwavelength 2 deg$^2$ COSMOS survey \citep{Scoville.etal.2007a}, the largest contiguous mosaic ever obtained using HST/ACS \citep{Koekemoer.etal.2007}. The zCOSMOS-bright sample is a purely flux limited part of the survey, selected at $I_{AB}<22.5$. It covers the central $\sim$1.7 deg$^2$ of the COSMOS field. This corresponds to the comoving size ($\sqrt{Area}$) of about 6.7, 25.1 and 54.3 \hh Mpc at redshift 0.1, 0.4 and 1, respectively, using the assumed cosmology. The zCOSMOS is targeted to obtain $\sim$ 20000 spectra of galaxies up to $z<1.4$ with sampling expected to reach a rather uniform $60-70\%$. At the moment, we have available spectra, and for most of them also possible various products, such as redshifts, luminosities, stellar masses, spectral line measurements etc. for 10644 objects, about half of the selected targets. This sample covers a slightly smaller area, of about 1.52 deg$^2$. The spatial sampling of these objects is very inhomogeneous, with $\sim 30\%$ on average. We refer to the catalogue containing only objects with reliable redshift measurements as the ``10k sample'' of the zCOSMOS survey.

We build also the so called ``30k sample'' of galaxies in the zCOSMOS survey area which satisfies the magnitude selection criteria $I_{AB}<22.5$, but for which spectroscopic measurement is not yet available or is not of sufficient reliability. For these galaxies we measure their photometric redshift probability function $P(z)$ using the ZEBRA code \citep{Feldmann.etal.2006}. The maximum in $P(z)$ we simply refer to as the photometric redshift. The uncertainty in the photometric redshifts which we use is estimated to be $\delta z = 0.023(1+z)$ \citep{Oesch.etal.inprep}.

The ZEBRA code has been employed also to calculate the rest-frame magnitudes of all $I_{AB}<22.5$ targets. The magnitudes are obtained as the best fit to the empirical set of spectral energy distribution (SED) templates normalised to each galaxy photometry \citep{Capak.etal.2007} at the best available redshift (spectroscopic or photometric). The stellar masses are obtained by fitting stellar population synthesis models to the SED of the observed magnitudes \citep{Bolzonella.etal.inprep, Pozzetti.etal.inprep}.

\subsection{Density field reconstruction}
\label{sub_densrec}

For the various COSMOS projects, environment has been characterised in a few different ways. The full volume density field reconstruction has been carried out by \citet{Scoville.etal.2007b}, \citet{Massey.etal.2007}  and \citet{Kovac.etal.2009}. \citet{Scoville.etal.2007b} reconstruct the galaxy large scale structure at $z<1.1$ using photometric redshifts of galaxies down to $I_{AB}<25$. Based on the observed shear field, \citet{Massey.etal.2007} produce maps of the large scale distribution of dark matter, resolved in both angle and depth, up to $z=1$. \citet{Kovac.etal.2009} describe the reconstruction of the density field in the zCOSMOS volume up to $z=1$, where the main ingredients are the high quality spectroscopic redshifts (with the uncertainty of about 100 \kms) of galaxies with $I_{AB}<22.5$ or subsamples of those galaxies. Moreover, \citet{Kovac.etal.2009} present an overview of the all estimators of the continuous environment  applied to the 10k zCOSMOS sample and discuss the importance of the scientific application on the exact choice of the density reconstruction method. In the following text, we will summarise the main steps in the zCOSMOS density field reconstruction method and define the galaxy samples used for this process. We refer to \citet{Kovac.etal.2009} for all the details.

Briefly, the density $\rho({\bf r})$ at a given point in space ${\bf r}(RA,DEC,z)$, can be defined as:

\be \rho({\bf r}) = \Sigma_i \frac{m_i W(|{\bf r}-{\bf r_i}|;R)}{\phi({\bf r_i})}. \label{eq_rhodef} \ee

\noindent
where $m_i$ is a weight related to some astrophysical property of an object (e.g. mass), $W(|{\bf r}-{\bf r_i}|;R)$ is a spatial smoothing function, and  $\phi(\bf r_i)$ is a function correcting the observed sample of objects to the complete sample of the same type of objects. The summation in Equation~\ref{eq_rhodef} goes over the observed galaxies which are used to reconstruct the density field. We call them tracer galaxies. In \citet{Kovac.etal.2009} we discuss thoroughly the possible values/functional forms of $m_i$, $W(|{\bf r}-{\bf r_i}|;R)$ and $\phi(\bf r_i)$ and a choice of tracer galaxies and their influence on the reconstructed density field.

We developed a new technique (ZADE) which allows us to incorporate both galaxies with spectroscopic and photometric redshifts in the density field reconstruction \citep{Kovac.etal.2009}. This approach is based on modifying the $P(z)$ function of galaxies from the 30k sample based on the proximity of galaxies from the 10k sample. We use counts of objects from the 10k sample in spheres of $R_{ZADE}$ along redshift to redistribute the ZEBRA calculated $P(z)$ into a new, ZADE-modified, $P_{ZADE}(z)$. We have done extensive tests on mock catalogues to justify this method. For the zCOSMOS survey, a density field reconstruction with $R_{ZADE} = 5$ \hh Mpc produces a density field without any major systematic error in the range $0.1<z<1$ (see the following Section~\ref{sub_recerr} and also Figure 4 in \citealt{Kovac.etal.2009}).

With the ZADE method, we use {\it all} $I_{AB}<22.5$ galaxies in the zCOSMOS area to reconstruct the density field, albeit with different quality of their redshift measurement. When calculating densities of the tracer galaxies, tracers with spectroscopic redshift are counted as one (i.e. $\delta$-function) at their measured redshifts, and the tracers with only photometric redshifts are counted as fractional objects according to $P_{ZADE}(z)$ value at the redshift of consideration $z$ (see also equation~\ref{eq_wrth} below). Formally, the ZADE approach is equivalent to $\phi({\bf r_i}) = 1$ for every tracer galaxy. This means that in the ZADE approach the mean intergalaxy separation is a characteristic separation of the total population of that sample of tracer galaxies, and not only of a part of that population with the measured spectroscopic redshift. The mean intergalaxy separation in the ZADE approach will therefore always be smaller than in some incomplete set of tracers, which is important in suppressing the shot noise effects.

In our reconstruction method we assume that the structures traced by the 30k sample of tracer galaxies are delineated by the 10k sample of tracer galaxies (the sample with reliable spectroscopic redshifts), i.e. that the 30k and 10k samples are equally biased tracers of the underlying matter distribution. Based on the selection criteria and observed distributions of the zCOSMOS galaxies,  this assumption is valid up to $z=1$ at least. We have done a number of careful checks on mock catalogues to validate this point (see Section~\ref{sub_recerr}).

For the  biasing analysis in this paper,  the density of galaxies
needs to match  exactly the method with which  the density of
matter is estimated. Here, we count
objects     within  a   spherical     top-hat    filter $W(|{\bf r}-{\bf r_i}|;R)$ with  smoothing scale $R=R_{TH}$,     following

\be
\label{eq_wrth}
W(\left| {\bf r}-{\bf r_i} \right|;R) = \left \{
                                  \begin{array}{lr}
                                  \frac{1}{\frac{4}{3} \pi R_{TH}^3} P_{i,ZADE}(z_i) &  \it{if} \  \left|{\bf r}-{\bf r_i} \right| \le R_{TH}   \\

                                  0 & \it{otherwise},
                                  \end{array}
\                                 \right.
\ee

\noindent
In the equation above, $|{\bf r}-{\bf r_i}|$ is the three dimensional distance between a tracer
galaxy and the point where the density is estimated.

We express the reconstructed density field in terms of the dimensionless density contrast $\delta(\bf{r})$ (commonly referred to also as ``overdensity'') defined as

\be 
\delta({\bf{r}})=\frac{\rho({\bf{r}}) - \rho_m(z)}{\rho_m(z)} ~. 
\ee

\noindent
Here $\rho_m(z)$ is the mean density at redshift $z$, which we calculate as the volume average. For a point {\bf r} in which a part of the cell defined by the smoothing scale $R$ of a filter $W$ falls outside of the survey limit, we apply an edge-correction by scaling the measured density with the fraction $f_e$ of the cell which is inside of the geometrical limits of the survey, such that the corrected density is $\rho/f_e$. For the analysis, we use only the cells for which at least half of the volume is within the survey limits.

The basic samples of all possible tracer galaxies (i.e. tracers selected only by $I_{\rm AB}  <$ 22.5), with both spectroscopic and photometric redshifts, are equivalent to those described in \citet[Section 4.2.1]{Kovac.etal.2009}. For the density field reconstruction we define four types of tracer galaxies:  the ``flux limited sample'' of galaxies with $I_{\rm AB}  <$ 22.5 in $0.1<z<1$ and three ``luminosity complete samples'' of galaxies satisfying the criteria $M_B<-19.5-z$ in $0.1<z<0.7$, $M_B<-20-z$ in $0.1<z<0.9$ and $M_B<-20.5-z$ in $0.1<z<1$. We include the passive evolution of $\Delta M_B = - \Delta z$ in selection of the luminosity complete samples in order to keep similar galaxies  in a unique sample at every redshift. The samples are selected to be complete for galaxies of both red/early and blue/late types (see Figure 1 in \citealt{Kovac.etal.inprep}).

We estimate $\rho_m(z)$ for each of these samples by adding up contributions of the tracer galaxies at each redshift. The contribution of any tracer galaxy to the mean density at some $z$, up to $z_{max}$, is calculated as $\Delta V(z)/V_{max}$, where  $\Delta V(z)$ is the volume of the individual redshift bin and $V_{max}$ is the volume of the zCOSMOS survey in which a tracer galaxy of consideration can be detected, limited by $z_{max}$. We take the passive evolution into account. The K-correction for the individual galaxies is obtained from the ZEBRA code. For practical purposes redshift is quantified in bins each $\Delta z=0.002$ wide.

\begin{figure}
\includegraphics[width=0.45\textwidth]{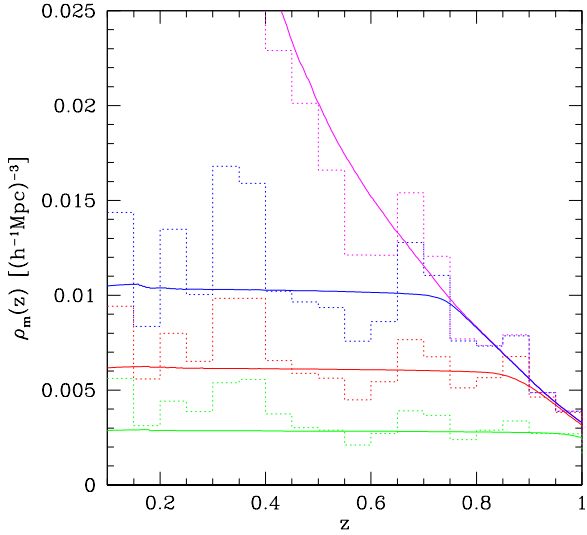}
\caption{\label{fig_voldens} Unity-weighted ($m_i=1$) mean volume density of the zCOSMOS tracer galaxies in $0.1<z<1$. The dotted lines are the mean volume densities of the zCOSMOS tracer galaxies obtained by dividing the number of galaxies with volume corresponding to redshift bins of 0.05. The continuous lines are smoothed mean densities, obtained by adding $\Delta V(z)/V_{max}$ contributions of each tracer galaxy in redshift bins 0.002 wide. The magenta, blue, red and green curves are for the flux limited, and $M_B<-19.5-z$, $M_B<-20-z$ and $M_B<-20.5-z$ luminosity complete samples, respectively. See text for more details.}
\end{figure}

The redshift distribution of the mean volume densities, weighted by unity, obtained by averaging the numbers of zCOSMOS galaxies in 0.05 redshift bins and the corresponding smooth $\rho_m(z)$ functions are presented in Figure~\ref{fig_voldens}. From this figure it is obvious that even the survey of the area of the zCOSMOS is dominated by inhomogeneities in the redshift distribution of galaxies of $\Delta z \sim 0.1$. On the other hand, the procedure to calculate $\rho_m(z)$ described above smooths the peaky distribution of the observed tracer galaxies, producing the mean volume density of luminosity complete tracer galaxies almost constant with redshift. The  mean intergalaxy separation $l$ of the zCOSMOS galaxies based on $\rho_m(z)$ is $\sim$4.6, $\sim$5.5 and $\sim$7.1 \hh Mpc for the $M_B<-19.5-z$, $M_B<-20-z$ and $M_B<-20.5-z$ samples, respectively. For the flux limited sample of tracer galaxies, $l$ increases from $\sim$1.9 \hh Mpc at $z=0.1$ to $\sim$3.3 \hh Mpc at $z=0.4$ and to $\sim$6.7 \hh Mpc at $z=1$.

\section{Measuring the biasing from the observational data}
\label{sub_method}

A number of methods to measure the nonlinear biasing from the observed
data  have been  proposed.  \citet{Szapudi.1998}  suggests to  use the
cumulant  correlators  of the  observed  distribution  of galaxies  in
redshift        surveys;        \citet{Matarrese.etal.1997}        and
\citet{Verde.etal.1998}  derive the first  two Taylor  coefficients of
the biasing function from the  bispectrum of galaxies from the survey.
SBD  proposed a  method, tested  on simulations,  to measure  the mean
biasing function relating the cumulative distribution functions (CDFs) of the
density fluctuations of galaxies and mass.

In this  work, we  follow the  method developed by  SBD to  derive the
mean biasing   function  $b(\delta)$ (or more specifically the conditional mean function $\langle \delta_g|\delta \rangle$) and  its second  moments  using   the  10k
zCOSMOS-bright  survey over  the redshift  interval 0.4-1.0.  Here, we
summarise the main points and limitations of the SBD method and refer the
interested readers  to the original paper  for all the  details.

We  denote  the CDFs of  density
fluctuations  of galaxies  and mass  $C_g(\delta_g)$  and $C(\delta)$,
respectively. Both distributions are functions of scale and
redshift   at   which   the   density  contrast   is   estimated   and
$C_g(\delta_g)$ also  depends on the  type of tracer galaxies  used to
reconstruct the overdensity field.  Assuming that the biasing relation
between galaxies and mass  is deterministic and monotonic, the biasing
function  could be  estimated from  the inverse  CDF
 of galaxies  at a given value (percentile)  of the CDF
 of mass (SBD):

\be  \delta_g (\delta) = C_g^{-1}[C(\delta)] , \label{eqCC} \ee

\noindent  where $C_g^{-1}[...]$  denotes the  inverse CDF  of galaxies.
SBD used simulations to show that equation~\ref{eqCC} can successfully
reproduce  the true conditional mean function $\langle \delta_g|\delta \rangle$  over  the  broad  range of  $\delta$
probed, overestimating it slightly in
the most dense regions. They  also conclude that the assumption of the
monotonicity  of  $\langle   \delta_g|\delta  \rangle$  is  valid  (or
represents a  very good approximation) on  scales of a  few Mpc, which
are the scales over which the biasing is measured in practice.

\subsection{PDF of density contrast of galaxies and mass}
\label{sub_pdf}

Current galaxy redshift surveys  provide data to calculate  the galaxy density contrast
and its PDF with sufficient accuracy up  to redshifts  of about
1.5 \citep[e.g.][]{Marinoni.etal.2005}.  Obtaining  the  mass  density  contrast observationally is on the other hand an extremely
difficult task. Direct reconstruction has been possible using galaxy peculiar velocities as tracers of mass fluctuations so far only in the local  universe \citep[out to
$\sim$  100 \hh Mpc;][]{Dekel.etal.1990, Dekel.etal.1999, Dekel.2000} or using weak lensing up to $z \sim 1$ \citep{Massey.etal.2007}, although with much lower resolution in the redshift dimension than is obtained for the galaxy density field from the galaxy redshift surveys. From a theoretical point of view it has been shown that in comoving space, the density contrast $\delta$ of matter follows, to a good approximation, a lognormal distribution $p(\delta)$ \citep{Coles&Jones.1991}. In this paper we will use this theoretical approximation of the matter distribution and in the following text, we will summarise the theoretical development necessary to calculate the PDF of the matter. However, we want to mention that there is a project in development to use simultaneously galaxy and matter density field (the latter one reconstructed from the weak lensing shear maps) in the COSMOS volume to estimate bias directly from the observed data.

The lognormal distribution of the matter fluctuations can be expressed as

\be  
p( \delta)_R = \frac{ (2 \pi \omega_R^2) ^{-1/2} } {1+ \delta} \exp \left\{ - \frac{[ \ln(1+ \delta)+ \omega_R^2/2]^2}{2 \omega_R^2} \right\} .   
\label{eq_logn}
\ee

\noindent
In the last equation, the parameter $\omega_R^2$ is defined as

\be   \omega_R^2 = \ln[1 + \langle \delta^2 \rangle_R]. \ee

\noindent
where $\delta$ is directly related to the variance $\sigma_R^2(z)$ of the density contrast field at redshift $z$ via

\be   \langle \delta^2 \rangle_R =  \sigma_R^2(z)  \ee

\noindent 
given that the density contrast field has zero mean. In the equations above, the index $R$ denotes the smoothing scale at which the density field is reconstructed. The value of $\sigma_R$ on a given scale is determined by the adopted cosmology. For the smoothing scales $R$ which are large enough to be in the  linear regime, its evolution with redshift can be modelled as

\be \sigma_R(z) = \sigma_R(z=0)D(z) \ee

\noindent
where $D(z)$ is the linear growth factor of density fluctuations, normalised to unity at $z=0$. For smaller scales, one would need nonlinear corrections.

Given that the derived lognormal form of the PDF of the matter overdensity field is calculated in real (comoving) space, and that the PDFs from surveys are obtained in redshift space, one has to convert both functions to the same space. If the redshift distortion affects both galaxies and matter in a similar way, we can expect that the mean biasing function $b(\delta)$ and the conditional mean function $\langle \delta_g|\delta \rangle$ in $z$ space will be similar to the one in real space:

\be  \langle \delta_{g,z}|\delta_z=\delta \rangle = \langle \delta_g|\delta \rangle. \ee

\noindent
SBD  show that the PDF (or CDF) of the mass density contrast in redshift space can also be well described with the lognormal function (equation~\ref{eq_logn}), with standard deviation obtained in redshift space. The relation between the azimuthally averaged standard deviations of mass fluctuations in real $\sigma_R(z)$ and redshift $\sigma_R^z(z)$ comoving space at high redshift is (e.g. SBD)

\be  \sigma_R^z(z) = [1+ \frac{2}{3}f(z) + \frac{1}{5}f^2(z)]^{1/2} \sigma_R(z) \ee

\noindent
based on the expression derived by \citet{Kaiser.1987}. For the calculations, we use the relation between the growth rate $f$ and the growth factor $D$ given by 

\begin{eqnarray}
f(\Omega_m(z), \Omega_{\Lambda}(z)) = \frac{d\ln D}{d\ln a} = \nonumber \\ 
 = -1 - \frac{\Omega_m(z)}{2} + \Omega_{\Lambda}(z) + \frac{5 \Omega_m(z)}{2g(z)}
\end{eqnarray}

\noindent
where $a=(1+z)^{-1}$ and

\be    D(z) = \frac{g(z)}{g(0)(1+z)}  \ee

\noindent
for linear fluctuations \citep{Carroll.etal.1992} and

\begin{eqnarray}
&& g(z) \approx \frac{5}{2}\Omega_m(z) \times \nonumber \\
&& \times \left[ \Omega_m^{4/7}(z) - \Omega_{\Lambda}(z) +  
\left(1 + \frac{\Omega_m(z)}{2}\right)\left(1 + \frac{\Omega_{\Lambda}(z)}{70}\right) \right]^{-1} ~. 
\end{eqnarray}

In the literature, there are other approximations available for the growth rate $f$ at different redshifts. For example, \citet{Lahav.etal.1991} derived $f \approx \Omega_m(z)^{0.6}$ and \citet{Wang&Steinhardt.1998} derived $f \approx \Omega_m(z)^{0.55}$, the latter one allowing the possibility that the current accelerated phase of the universe is due to quintessence (a time-evolving, spatially inhomogeneous component with negative pressure) and not due to the cosmological constant.

\section{Errors in the mean biasing function and biasing parameters}
\label{sec_tests}

The mean  biasing function (or the conditional mean function $\langle \delta_g|\delta \rangle$) and its moments  derived using  the method
described  in  Section~\ref{sub_method} will
contain uncertainties  due to the finite volume of  the survey  (cosmic variance errors) and
 the use of discrete tracers (i.e. galaxies) to reconstruct the continuous density field (shot noise errors).  In addition to  this, the
results   on  the  biasing   will  contain   errors  related   to  the
reconstruction of the density field.   We use mock catalogues to asses
the contribution  of each of these errors.

The mock catalogues which we employ are based on the lightcones for the COSMOS survey kindly provided by \citet{Kitzbichler&White.2007}. The mock catalogues are built from the N-body Millennium Simulation \citep{Springel.etal.2005} using the semi-analytic modelling of galaxy properties based on \citet{deLucia&Blaizot.2007} with some modifications described in \citet{Kitzbichler&White.2007}. We cut the mock catalogues such that their area and redshift limits match the zCOSMOS survey. Depending  on the  errors which we  test, we
use mocks  with different flux  or luminosity limits and  of different
sampling.

We  reconstruct the  overdensity field  on a  grid with
separation  of 0.5  \hh  Mpc  in the $RA-DEC$  plane  and 0.002  in
the $z$ direction. As for the real data, we
use every  tracer galaxy to  reconstruct the density field, using either its spectroscopic redshift or its ZADE-modified artificial photometric redshift probability distribution. As we have mentioned previously, this is equivalent to
$\phi_i=1$   for   every   tracer   galaxy   following   notation  of
equation~\ref{eq_rhodef}. We   limit
ourselves    to the  use    of   the   unity-weighted   overdensities,
i.e.  $m_i=1$ in the same equation. The  mean density  in the  individual mock  catalogues is estimated by smoothing the $N(z)$ distributions for the flux limited samples. For  the luminosity complete samples the mean density is estimated  for each redshift bin  simply as the number of objects divided by the volume.  While  for the  biasing analysis  we  use
redshift  bins of  0.3  width, to
estimate the  mean number  of objects we  use redshift  bins  of 0.4. These bins are broadened symmetrically by $\delta z = 0.05$.

To be consistent with the analysis of the zCOSMOS data presented in Section~\ref{sec_analysis}, when calculating the mean
biasing function using the mocks (in order to evaluate various errors), we assume $\sigma_8=0.8$, roughly consistent with the latest WMAP results \citep{Dunkley.etal.2009}, even though
the mock catalogues are based on the simulations with $\sigma_8=0.9$. Similar to recent works,  we  choose the
logarithmic representation of overdensities  in order to emphasise the
behaviour in the underdense regions. For simplicity, we will refer to the conditional mean function of $\log(1+\delta_g)$ at a given $\log(1+\delta)$ also as $\langle \delta_g|\delta \rangle$.

Before  we proceed  with  measuring  the errors  in the conditional mean function $\langle \delta_g|\delta \rangle$ 
function  and the second  moments of the mean biasing function,  we  need  to estimate  the  scale $R$
(smoothing top-hat window) at which we can reconstruct the overdensity
field  without serious systematic  errors.  We  reconstruct  the  galaxy
overdensity  field on the so called 10k+30kZADE and  40k mock  catalogues. 
Both of these types of mock catalogue have the same geometrical constrains as the zCOSMOS survey. In the 40k mock catalogue all galaxies have measured spectroscopic redshifts. In the 10k+30kZADE catalogue, only $\sim 10000$ galaxies with $I_{AB} < 22.5$ have known spectroscopic redshifts. Their $RA-DEC$
distribution has been matched to the complicated sampling (see Figure 4 in \citealt{Lilly.etal.2009}) and the redshift success rate of the 10k zCOSMOS sample.  We take  into account $I_{AB}<22.5$ objects  without
redshift by defining their photometric redshift probability distribution $P(z)$. In the mock catalogues, we model the initial $P(z)$ to be a Gaussian distribution with $\sigma=0.023(1+z)$, randomly offset in redshift using offsets selected from the same distribution. For the density field reconstruction this probability is then modified using the ZADE algorithm to yield $P_{ZADE}(z)$ (see Subsection~\ref{sub_densrec}). For computational purposes, $P_{ZADE}(z)$ is discretised in redshift bins of 0.002.
  
\begin{figure}
\centering
\includegraphics[width=0.5\textwidth]{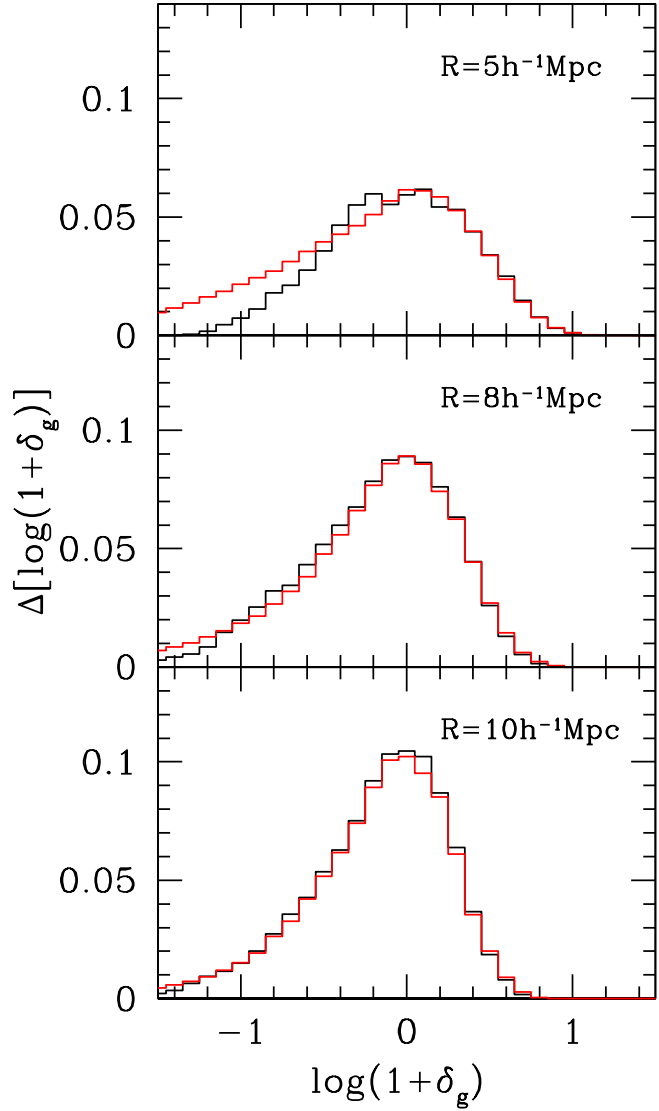}
\caption{\label{fig_pdfmock} Example of the PDF of the galaxy density contrast field using a mock catalogue. The density field is reconstructed using the top-hat smoothing filter of 5, 8 and 10 \hh Mpc going from the top to the bottom panel. The black histogram corresponds to the reconstruction with the 40k catalogue (every galaxy has a measured spectroscopic redshift) and the red histogram corresponds to the reconstruction with the 10k+30kZADE catalogue (10k-like sample of galaxies with a measured spectroscopic redshift, the rest of the $I_{AB}<22.5$ galaxies have the ZADE-modified photometric redshift). Binning is carried out in $\log(1+\delta_g)$ units. At the current status of the zCOSMOS survey, we need scales of at least 8 \hh Mpc to reconstruct the density field for the biasing analysis with acceptable errors at every $\delta_g$ up to $z \sim 1$.}
\end{figure}

To test the minimum smoothing scale, we use $I_{AB}<22.5$ samples of galaxies and three smoothing scales: 5,  8 and 10 \hh Mpc. As an example, the  PDFs of the reconstructed
density  contrast for the 40k and 10k+30kZADE samples is  presented in  Figure~\ref{fig_pdfmock} for one of the mock catalogues. For the
current status of the zCOSMOS survey  we need scales of at least 8 \hh
Mpc to reconstruct the overdensity field with acceptable errors at every $\delta_g$ up to $z \sim 1$.

\subsection{Cosmic variance}

One  of the errors  entering the  biasing analysis  occurs due  to the
relatively small  survey volume and consequent noise in $C_g(\delta_g)$. This  error is usually  termed cosmic
variance. To quantify it, we  employ 12 mock catalogues designed to be
equivalent in  the terms of geometrical  and magnitude selection  constrains to
the zCOSMOS survey.  We use the full 40k catalogues to select the different subsamples of tracer galaxies to reconstruct the
density field.

\begin{figure}
\includegraphics[width=0.45\textwidth]{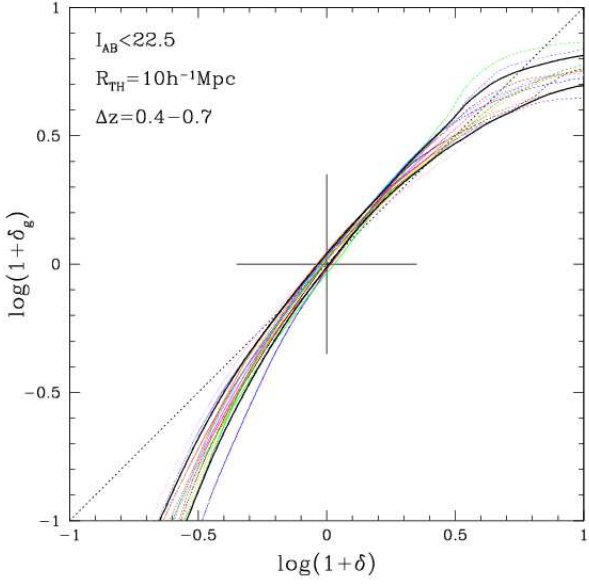}
\caption{\label{fig_cosvar}Effect of the cosmic variance errors on the conditional mean function $\langle \delta_g|\delta \rangle$. At every $\delta$ value, cosmic variance is causing a spread in the corresponding $\delta_g$ values. The mean $\langle \delta_g|\delta \rangle$ functions are obtained from the 12 mock catalogues, plotted as the thin dotted lines, where the mock galaxy density field has been reconstructed with the flux limited sample of galaxies and top-hat filter of 10 \hh Mpc in $0.4<z<0.7$. In  all figures
containing the $\langle \delta_g|\delta \rangle$ function,  starting with this,  we mark
the case  of a  no-biasing $b_{L}=1$ with  the black dotted  line. The
cross in the  middle of the panels is drawn for  a reference and marks
the $\delta_g=\delta=0$ case. The standard deviation of $\delta_g$ values in the mocks at every $\delta$ are plotted with the thick continuous lines, centred at the mean $\delta_g$. This is the effective cosmic variance noise expected in a single reconstruction, i.e. in the actual data.}  
\end{figure}

As an example, we show  in Figure~\ref{fig_cosvar} the conditional mean functions $\langle \delta_g|\delta \rangle$ obtained for the 12 mock catalogues, where the mock galaxy density field has been reconstructed with the flux limited sample of galaxies and top-hat filter of 10 \hh Mpc in $0.4<z<0.7$. Due to cosmic variance, there is a dispersion of $\delta_g$ values corresponding to every $\delta$ value in the biasing analysis since every mock will uniquely map $\delta$ to a single (mean) $\delta_g$ value. In the mocks, the measured range of possible $\delta_g$ values is largest in the most underdense regions, where the standard deviation $\sigma$ of $\log(1 + \delta_g)$ values is $\sigma \sim 0.1$. The $\delta_g$-dispersion gets smaller towards regions near the mean density, where  $\sigma < 0.05$. In the most overdense regions, the cosmic variance error is increasing again, up to $\sigma \sim 0.05$.  We  will use  the standard deviation in the conditional mean function $\langle \delta_g|\delta \rangle$  and
in the biasing parameters obtained from the 40k mock catalogues as the  errors in corresponding values derived from
the zCOSMOS sample.

It is important to point out that the the conditional mean function $\langle \delta_g|\delta \rangle$ in each of the mock catalogues is derived with respect to the same theoretical underlying matter PDF, as it would be the case with the observations. To obtain the ``true'' cosmic variance, one would need to derive $\langle \delta_g|\delta \rangle$ with respect to the PDF of the matter using the DM particles in each of the mock catalogues (which we do not have). Moreover, stochasticity may also contribute to the measured scatter in the $\langle \delta_g|\delta \rangle$ function.

\subsection{Shot noise errors}
\label{sub_shotnoise}

Sparse sampling  artificially enhances both the
positive and the negative tails  of the density contrast distribution. This
effect broadens  the PDF  and thus steepens the conditional mean function $\langle \delta_g|\delta \rangle$ (e.g. SBD). To  evaluate  the shot  noise errors, we first use the 12 mocks including the full geometrical  constrains of the zCOSMOS
survey and limited by the rest-frame magnitude $M_B<-18$ (ignoring the
luminosity  evolution  for the  moment). This  is the faintest magnitude  cut for which  the given mock
surveys $r \le$ 26 \citep{Kitzbichler&White.2007} are luminosity complete
up to $z=1$. The mean separation $l$ between galaxies  is
about 2.7 \hh  Mpc in the individual mock  catalogues, which should be
sufficient to obtain the mean biasing function and its moments without
shot noise errors for scales of 8 \hh Mpc and larger. To quantify  the effect  of the shot  noise errors we then resample these
mock catalogues at random such that the mean distance between galaxies $l$
is 3, 4, 5,  6, 7 and 8 \hh Mpc. We  derive the conditional mean function $\langle \delta_g|\delta \rangle$ and the second moments of  the mean biasing function
using the overdensity  fields reconstructed
from  the resampled  mock  catalogues in  two  redshift bins:  $\Delta
z=0.4-0.7$ and  $\Delta z=0.7-1$.

\begin{figure}
\includegraphics[width=0.45\textwidth]{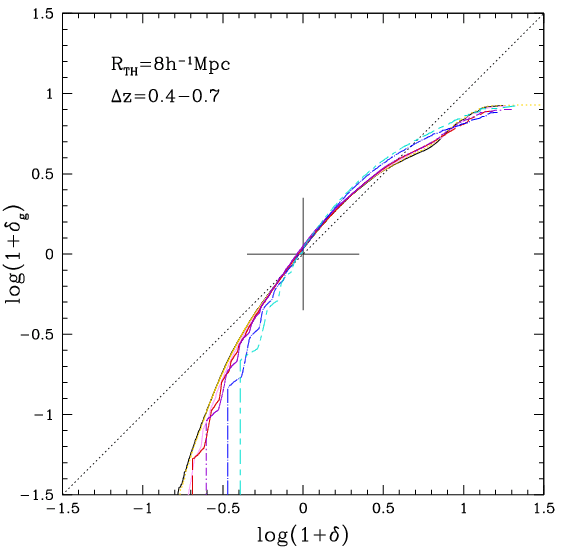}
\includegraphics[width=0.45\textwidth]{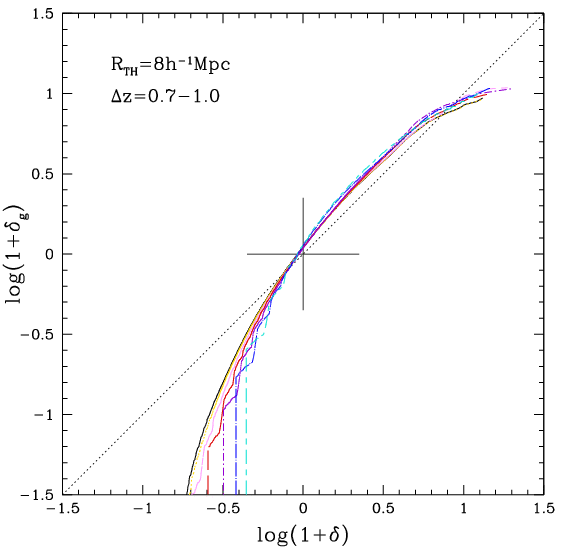}
\caption{\label{fig_shotnoiseexmpl}Effect  of galaxy sampling on the conditional mean function $\langle \delta_g|\delta \rangle$. The $\langle \delta_g|\delta \rangle$ function between galaxies ($M_B<-18$)
and mass  for a top-hat smoothing of  8 \hh Mpc in  two redshift bins:
$\Delta z =0.4-0.7$  and $\Delta z =0.7-1.0$ is  presented in the left
and right  panel, respectively. The $\langle \delta_g|\delta \rangle$ function with  the full
sampling in  the given mock catalogue ($l  \sim 2.7$ \hh  Mpc) is
presented with  the black continuous line.  The  various curves correspond
to the  different mean galaxy  separation $l$ of 3 (yellow), 4 (pink), 5
(red),  6 (violet),  7 (blue)  and 8  (cyan) \hh  Mpc.}
\end{figure}

The effect of  the sampling on the  $\langle \delta_g|\delta \rangle$ function derived from
the      zCOSMOS      size       survey      is      presented      in
Figure~\ref{fig_shotnoiseexmpl}  for one of  the mock  catalogues. The
shot noise errors can change the shape of $\langle \delta_g|\delta \rangle$  in
the underdense  regions dramatically, the effect  clearly depending on
the value of $l$ (relative to $R$), i.e. on then mean number of galaxies per sampling region. In the underdense regions, the value of the mass  density contrast associated with
a given galaxy  density contrast is artificially  shifted to
 higher  values when subsequently sampling  smaller fractions  of the
population  of tracer  galaxies.  This  is important  given  that this
minimum $\delta$ value, below which the galaxy density field
does  not trace  that  of the  matter, is  often interpreted  as the
minimum  mass density  contrast below  which the  formation  of tracer
galaxies  is partially  or  completely suppressed  \citep[e.g.][]{Somerville.etal.2001, Marinoni.etal.2005}.

\begin{figure}
\includegraphics[width=0.45\textwidth]{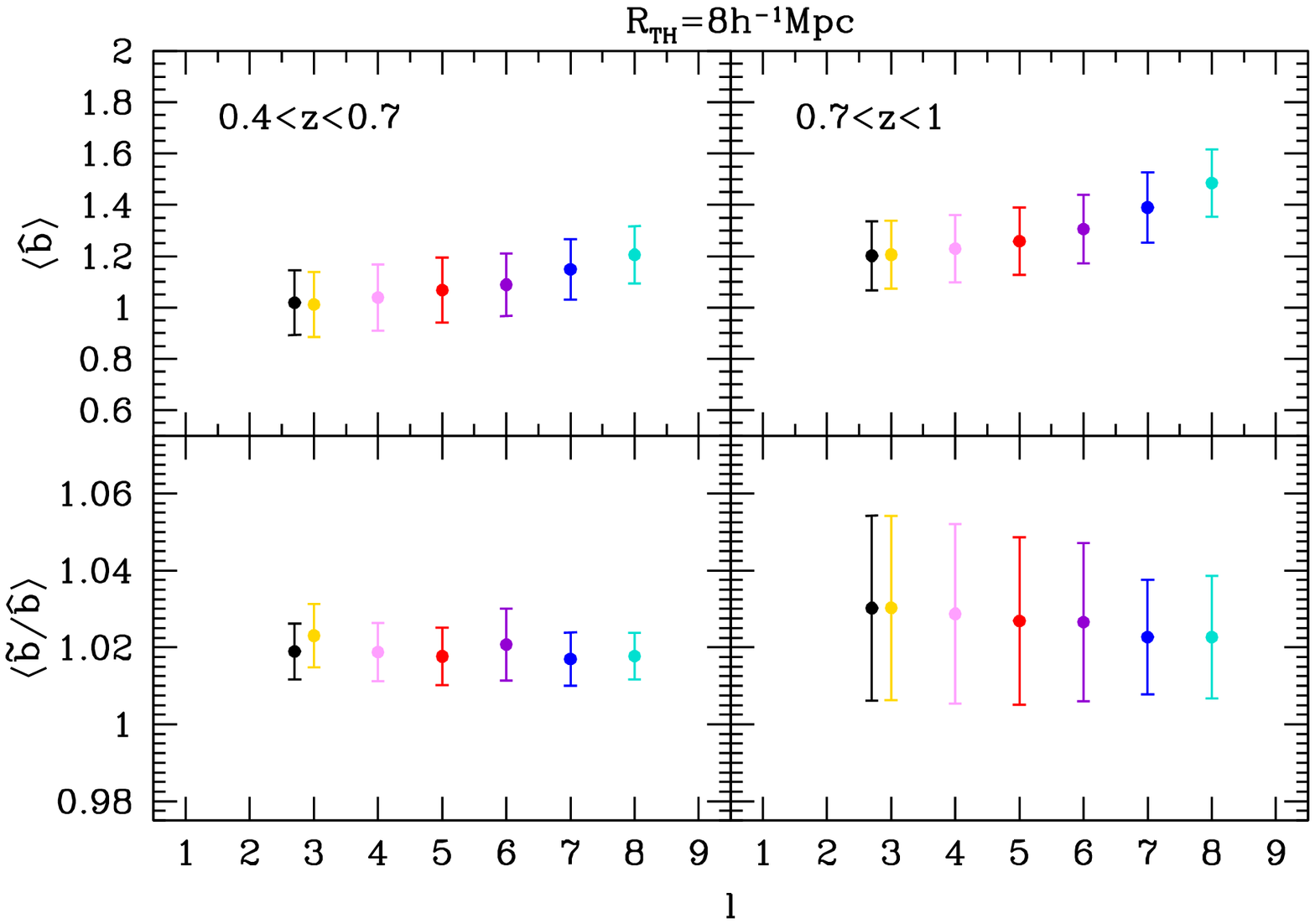}
\includegraphics[width=0.45\textwidth]{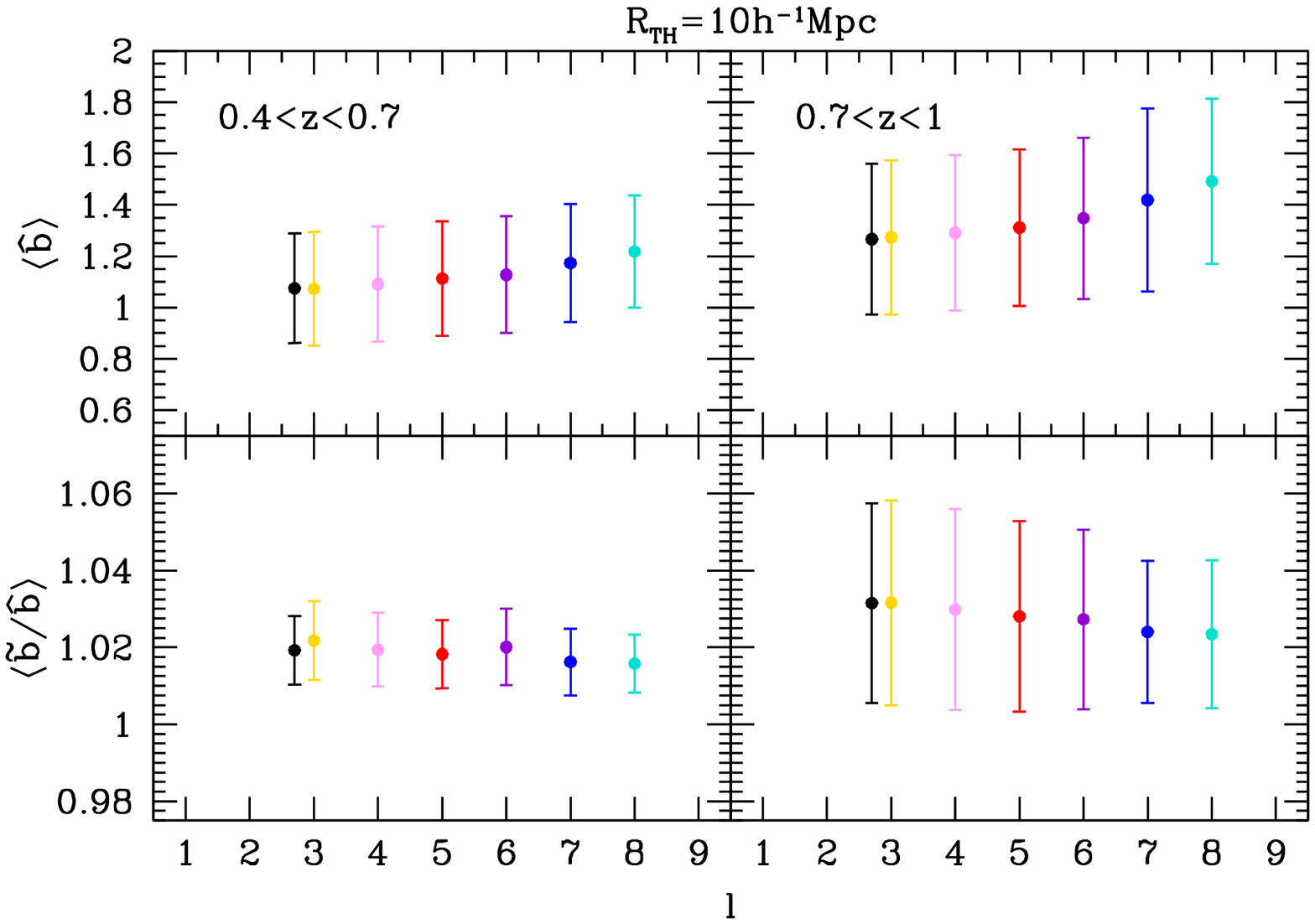}
\caption{\label{fig_shotnoisesummary}Summary of the effect of the sampling of galaxies on the second moments of the mean biasing function. Individual points correspond to the mean biasing parameters: linear bias $\hat{b}$ (top panels) and nonlinearity $\tilde{b}/\hat{b}$ (bottom panels) obtained by averaging results from 12 mocks, plotted as a function of the mean intergalaxy separation $l$. The vertical bars correspond to the standard deviation of the parameters from these 12 mocks. The colour coding is the same as in the previous figure. The four panel plots on the left and right side are obtained for the galaxy density field reconstructed on $R=8$ \hh Mpc and $R=10$ \hh Mpc, respectively. In each of the four panels, the left hand plots refer to $0.4<z<0.7$ and the right hand plots refer to $0.7<z<1$.}
\end{figure}

The  summary of  the effect  of the  shot noise errors  on  the second
moments   of    the   mean   biasing   function    is  shown   in
Figure~\ref{fig_shotnoisesummary}.  We average the  parameters $\hat{b}$
and $\tilde{b}/\hat{b}$  estimated from the 12  mocks and present  the averaged
parameters  with  the errors  calculated  as  the  rms values  of  the
corresponding parameters  in  these  12  mocks.  The linear  biasing  parameter  $\hat{b}$
increases constantly with increasing mean intergalaxy separation, while
there  is  some indication  that  the nonlinearity parameter  $\tilde{b}/\hat{b}$
slightly  decreases with increasing  mean intergalaxy  separation, at least in the higher redshift bin. For
example, for  the smoothing scale 8  \hh Mpc the linear  biasing parameter
increases by $\sim  20\%$ and nonlinearity decreases by  $\sim 0.6 \%$
when  changing  the  mean  intergalaxy  separation from  3  to  8  \hh
Mpc.  The  effect   of  the  shot noise  errors  on  the
nonlinearity parameter can  be practically neglected on smoothing scales  of 8 \hh Mpc and larger. Note that the presented standard deviations of the biasing moments from different mocks are dominated by the cosmic variance.

\subsection{Reconstruction errors}
\label{sub_recerr}

The reconstructed  galaxy density  field is meant to be the  density field  of the
full  population of tracer galaxies.  In  practice, the whole
population of galaxies is never available, and the measured properties
of  sample galaxies,  relevant  for the  reconstruction, contain  some
errors, such as errors in spectroscopic redshifts or ZADE-modified redshift photometric probability functions  in the samples  of  tracer  galaxies. We refer to reconstruction errors as those errors which arise due to the fact that only a part of the total population of tracer galaxies with their measured properties is used for the density field reconstruction instead of the full population with their true properties. Given that we use the ZADE approach, we can easily differentiate between the reconstruction and shot noise errors, as defined here. However, when using only a fraction of tracer galaxies with measured redshifts to reconstruct the density field (and correcting statistically for the galaxies without measured redshifts), it can be difficult to separate between such defined reconstruction and shot noise errors.

To gain an understanding of the effect of reconstruction errors in the 10k zCOSMOS density field on the biasing analysis, 
we derive the conditional mean function $\langle \delta_g|\delta \rangle$ and biasing parameters using
12 mock 10k+30kZADE catalogues, 
for the two samples of tracer galaxies: flux limited $I_{AB}<22.5$ and volume limited 
$M_B<-20-z$. We compare the results to the corresponding ones derived from the 40k mock catalogues. We  use only  the
smoothing  scale   of  8  \hh  Mpc,   given  that  the   error  in  the
density field reconstruction is  generally larger for smaller  smoothing scales (see
Section 5 in \citealt{Kovac.etal.2009}).

\begin{figure}
\includegraphics[width=0.23\textwidth]{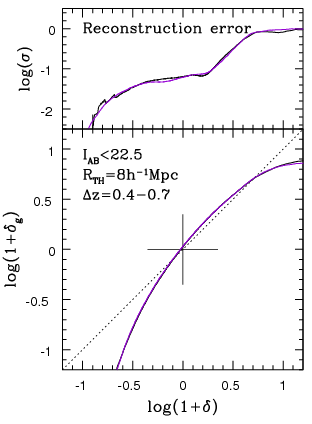}
\includegraphics[width=0.23\textwidth]{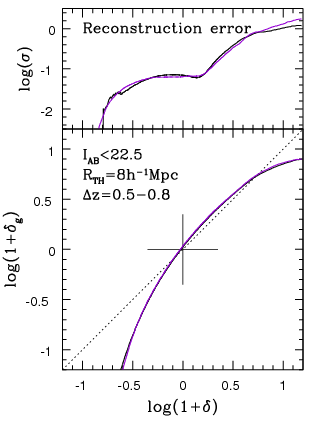}
\includegraphics[width=0.23\textwidth]{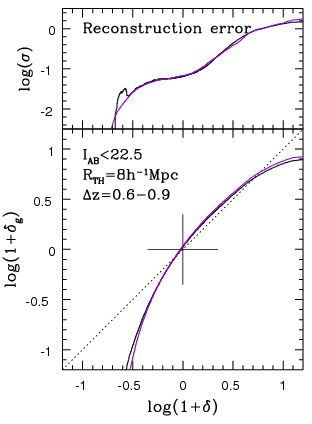}
\includegraphics[width=0.23\textwidth]{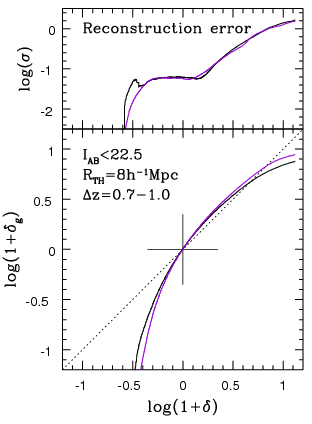}
\caption{\label{fig_meanbiasrecflux} Reconstruction errors in the mean biasing analysis for the flux complete samples. Lower panels: the $\langle \delta_g|\delta \rangle$ function of the overdensity field of galaxies ($I_{AB}<22.5$) for a top-hat smoothing of 8 \hh Mpc for the 10k+30kZADE reconstruction (violet) and 40k reconstruction (black). The curves are obtained by averaging results from 12 mock catalogues of the same type. Upper panels: the scatter in the corresponding $\langle \delta_g|\delta \rangle$ functions plotted below. The  scatter at fixed $\delta$ values is calculated as the standard deviation ($\sigma$) of $\delta_g$ values from 12 40k-type mocks (black) and from 12 10k+30kZADE-type mocks (violet).}
\end{figure}

\begin{figure}
\includegraphics[width=0.23\textwidth]{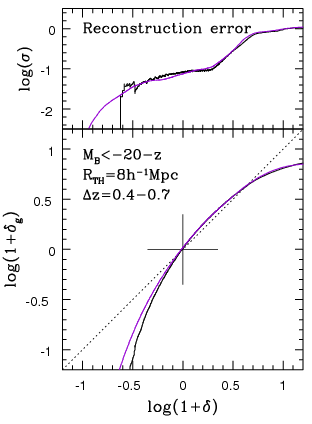}
\includegraphics[width=0.23\textwidth]{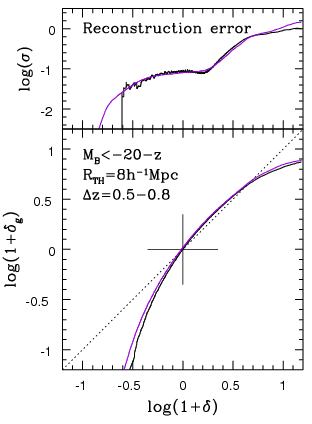}
\includegraphics[width=0.23\textwidth]{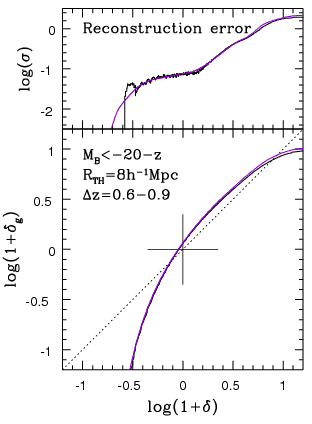}
\caption{\label{fig_meanbiasrecmB} Reconstruction errors in the mean biasing analysis for the luminosity complete samples. The curves have the same meaning as in Figure~\ref{fig_meanbiasrecflux}, but the zCOSMOS and mock overdensity field is reconstructed with the $M_B<-20-z$ galaxies.}
\end{figure}

The resulting $\langle \delta_g|\delta \rangle$ functions for the 10k+30kZADE (observed) and 40k (true) mocks
for smoothing top-hat scale of 8 \hh Mpc, obtained by averaging the
conditional mean functions $\langle \delta_g|\delta \rangle$ from the individual mock catalogues, are presented
in   the  lower   panels  of  Figures~\ref{fig_meanbiasrecflux} and~\ref{fig_meanbiasrecmB} for the  flux limited and $M_B<-20-z$ limited
samples of  galaxies, respectively. The error  in the reconstruction of the
density  field changes  the  shape  of $\langle \delta_g|\delta \rangle$  in
different  ways for  the flux  and luminosity complete samples of  tracer
galaxies, as following. The reconstruction errors for the flux limited sample of galaxies are negligible in low redshift bins, and they increase in higher redshift bins, artificially increasing local bias $b(\delta, z, R)$ values in the regions of the both highest and lowest density contrasts. For the volume limited samples the reconstruction error is manifested by lowering the local bias $b(\delta, z, R)$ values in the most underdense regions, and this error decreases with redshift. The different effect of the reconstruction error in the flux and luminosity complete samples arises from the fact that we are using the same ZADE-modified probability functions for the objects without spectroscopic redshift in both samples, but the overall numbers of galaxies in the flux and luminosity complete samples are different. These differences change with redshift (see Figure~\ref{fig_voldens}), and therefore the reconstruction error in the biasing analysis manifests itself differently in different redshift bins. In summary, the reconstruction error is most notable in the $\delta<0$ regions. At the reconstructed value of galaxy density field $\log(1+\delta_g)=-1$, the reconstruction error can cause differences up to $\sim \log(1+\delta)=0.1$. Any change in the conditional mean function $\langle \delta_g|\delta \rangle$ in the underdense region smaller than this value can therefore not be regarded as significant.

In the upper  panels of Figures~\ref{fig_meanbiasrecflux}   and~\ref{fig_meanbiasrecmB} we present the corresponding rms values ($\sigma$) of the average $\langle \delta_g|\delta \rangle$ value from  these  12 mocks, over the whole range of $\delta$ values. These $\sigma$ values correspond to the cosmic variance in the reconstructions of the galaxy density field with the 40k and 10k+30kZADE samples. For $\log(1+\delta) > -0.5$, $\sigma$ is almost identical for the 40k and 10k+30kZADE reconstructions for the both flux and luminosity complete samples. The fact that the conditional mean function $\langle \delta_g|\delta \rangle$ does not vanish at $\log(1+\delta)<-0.5$ for the 10k+30kZADE reconstruction with the luminosity complete samples is due to the ZADE-approach. Given that the matter distribution is lognormal, the total effect of this artificial ``filling'' of the underdense regions with ``fractional'' galaxies is negligible on the final results.

\begin{figure}
\includegraphics[width=0.45\textwidth]{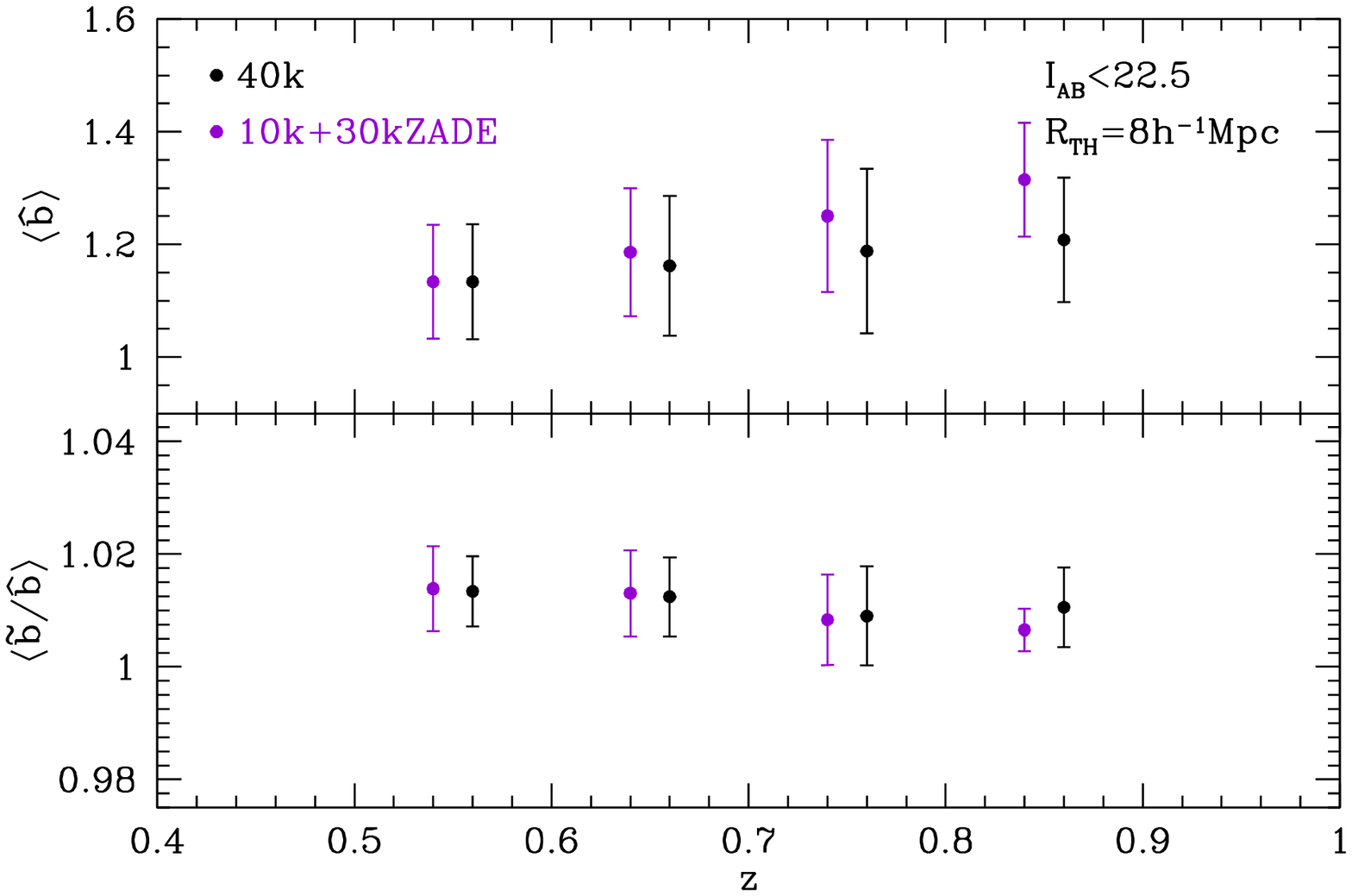}
\includegraphics[width=0.45\textwidth]{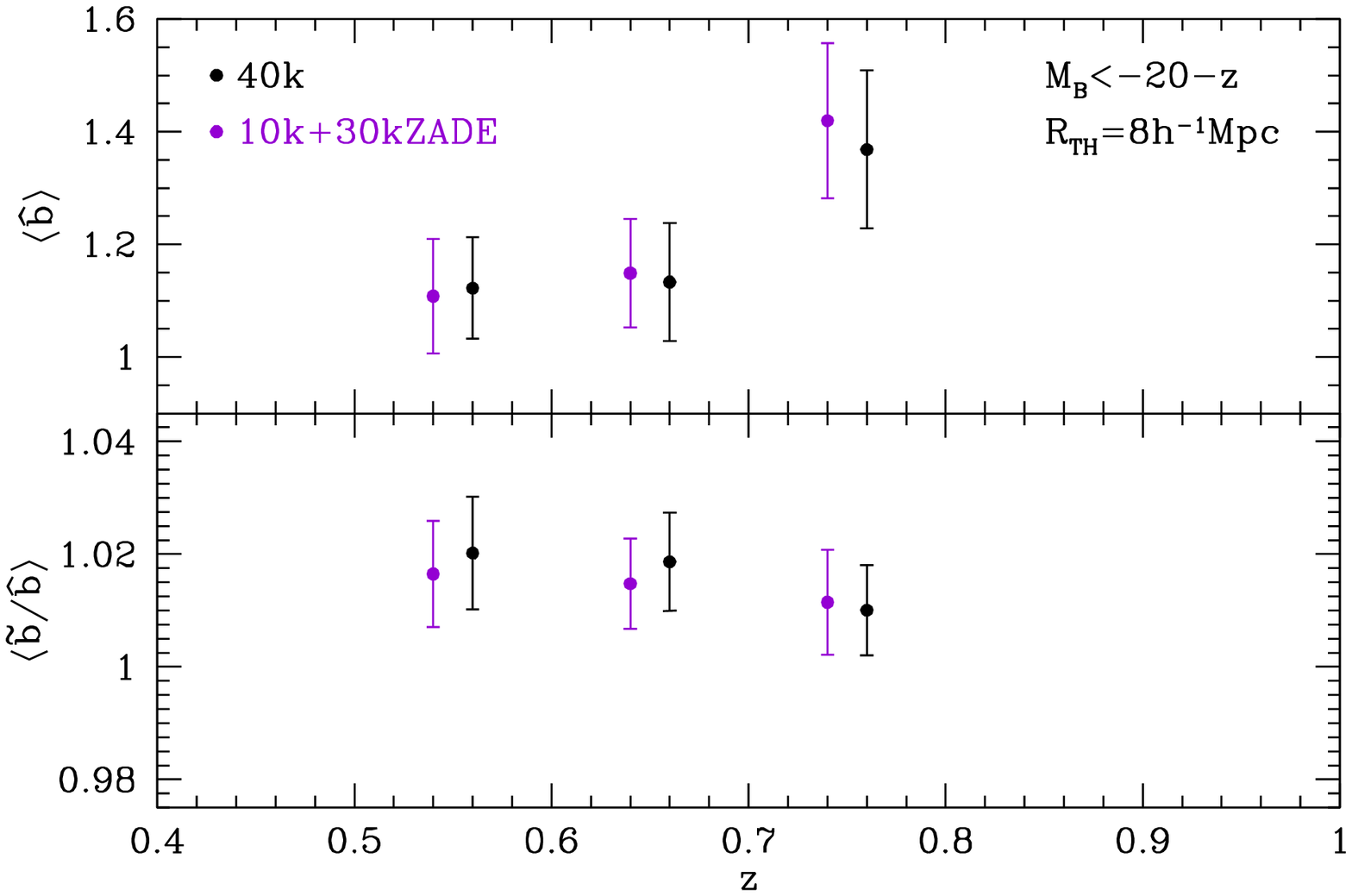}
\caption{\label{fig_biasparrec} Summary of the effect of the reconstruction errors on the second moments of the mean biasing function. Individual points correspond to the mean biasing parameters: $\hat{b}$ (top panels) and $\tilde{b}/\hat{b}$ (bottom panels) obtained by averaging results from 12 mocks. The vertical bars are the corresponding standard deviations.}
\end{figure}

In addition, we compare  the $\hat{b}$ and $\tilde{b}/\hat{b}$ parameters from
the      10k+30kZADE      and      40k  
    samples
(Figure~\ref{fig_biasparrec}).  The reconstruction error in  the $\hat{b}$ parameter
acts  as an artificial  increase of  this parameter for both flux limited and luminosity complete tracer galaxies.  On the
other  hand, we see that the nonlinearity  parameter $\tilde{b}/\hat{b}$  is almost
not affected by the error in the reconstruction.

\section{Biasing analysis: results and their interpretation}
\label{sec_analysis}

We derive the conditional mean function $\langle \delta_g|\delta \rangle$ and the biasing parameters $\hat{b}$ and $\tilde{b}/\hat{b}$ of the 10k zCOSMOS galaxies using
four galaxy samples  to reconstruct the density field and obtain the PDF of the density contrast: $I_{AB} <
22.5$ flux limited sample  and three luminosity complete samples of
$M_B<-19.5-z$, $M_B<-20-z$  and $M_B<-20.5-z$ galaxies.  We derive the
density  field  following  equation~\ref{eq_rhodef} with  the  top-hat
three   dimensional filter  (equation~\ref{eq_wrth}),   using
smoothing scales of 8, 10 and  12 \hh Mpc. The range of the smoothing scales is limited by the minimum of 8 \hh Mpc at which  we  can   reliably  reconstruct  the
overdensity field (based on the reconstruction
method  tested   on  the  mocks) and the transverse size of the zCOSMOS field. We use
the ZADE approach  to account for galaxies without reliably measured
 spectroscopic redshift  and therefore  $\phi_i=1$  for every
tracer galaxy. If not otherwise stated, we use the unity-weighted overdensity field ($m_i=1$) for the biasing analysis. The zCOSMOS selection catalogue is based on the 0.1 arcsec resolution HST images in F814W filter \citep{Koekemoer.etal.2007}, supplemented by photometry from a high resolution CFHT image in $i$ filter in the case when the HST data is missing or when the HST photometry is not of required quality. With this combination of underlying imaging only a negligible part of the zCOSMOS area is affected by the foreground stars \citep{Lilly.etal.2009}, and there is no need for masking of some regions for the density field reconstruction.

\begin{figure}
\includegraphics[width=0.24\textwidth]{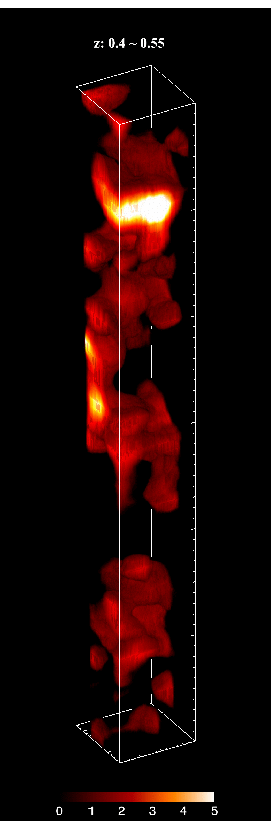}
\includegraphics[width=0.24\textwidth]{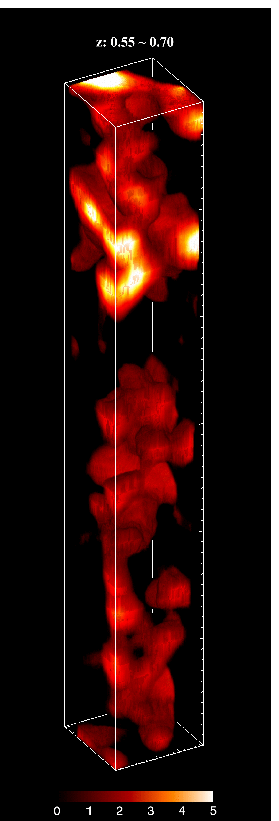}
\includegraphics[width=0.24\textwidth]{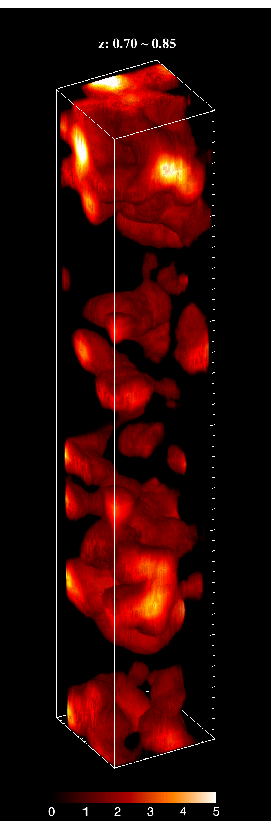}
\includegraphics[width=0.24\textwidth]{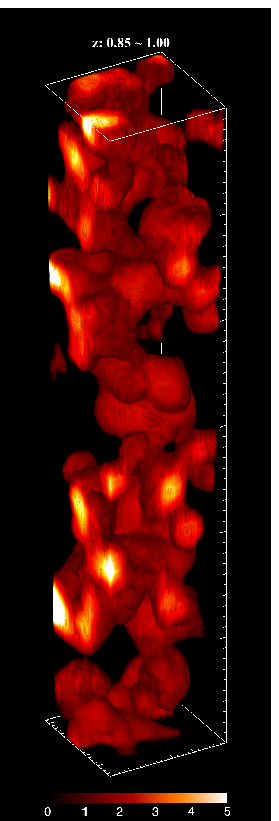}
\caption{\label{fig_densityfield}zCOSMOS overdensity field reconstructed with $R=8$ \hh Mpc and flux limited tracer galaxies in $0.4<z<1$. The colour scale on the bottom is given in the $1+\delta$ units. The horizontal axis are RA and DEC, the vertical axis is redshift. The size of the box is 35, 40, 45 and 50 \hh Mpc along RA and DEC axis from lower to higher redshift and $\sim 0.15$ along redshift. We plot only structures above the mean density ($1+\delta > 1$), in order to increase the visibility.} 
\end{figure}

As for the reconstruction of  galaxy overdensity field on the mocks that was used to obtain the contribution of various errors on the conditional mean function$\langle \delta_g|\delta \rangle$ 
and the second  moments of the mean biasing function (described above), we reconstruct the zCOSMOS galaxy overdensity  field on a grid with the points  in the $RA-DEC$ plane separated by 0.5 \hh  Mpc and with $\Delta z = 0.002$. The overdensity field reconstructed with $R=8$ \hh Mpc and flux limited tracers in $0.4<z<1$ is presented in Figure~\ref{fig_densityfield}. The complex, cosmic-web appearance of the density field, consisting of cluster-like structures,
surrounded by empty, void-like regions, is detected in the whole redshift range probed (see \citealt{Kovac.etal.2009} for the more detailed discussion of the structures in the zCOSMOS overdensity field).

We carry out the biasing analysis
in  the redshift  range $0.4<z<1$,  starting  at $z=0.4$  in order  to
exclude redshift slices in which for the majority of grid points the large fraction of filter aperture (i.e. larger than 0.5) falls outside of  the
survey volume.  We calculate the  density  contrast in  four  redshift  bins:
$0.4<z<0.7$, $0.5<z<0.8$, $0.6<z<0.9$ and $0.7<z<1$, which overlap  in order to suppress the effect
of the  cosmic variance. This is clearly visible in the zCOSMOS overdensity
field plots  (Figure~\ref{fig_densityfield}; see also Figures 12 and 17 in \citealt{Kovac.etal.2009}) and hampers the expected evolution in the distribution of large scale structures with cosmic time (see Figure 19 in \citealt{Kovac.etal.2009}). At lower redshift, the mean overdensity in the  individual redshift  slices differ from zero by a few percent. However, in $0.6<z<0.9$ and $0.7<z<1$, the mean density is about $0.1-0.15$ for all but $M_B<-20.5-z$ sample in $0.6<z<0.9$, for which the mean overdensity is about $0.2$. Therefore, the bias values for these samples needs to be taken with caution.

\begin{figure}
\includegraphics[width=0.45\textwidth]{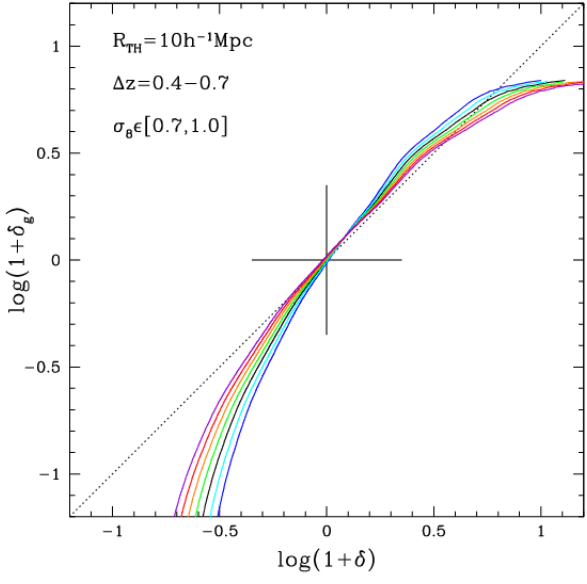}
\caption{\label{fig_biassigmaval} Conditional mean function $\langle \delta_g|\delta \rangle$ estimated for different $\sigma_8$ values. The parameter $\sigma_8$ increases from 0.7 to 1 in steps of 0.05. The resulting $\langle \delta_g|\delta \rangle$ function is presented in blue for $\sigma_8=0.7$, cyan for $\sigma_8=0.75$, black for $\sigma_8=0.8$, green for $\sigma_8=0.85$, orange for $\sigma_8=0.9$, red for $\sigma_8=0.95$ and violet for $\sigma_8=1$.}
\end{figure}

As discussed in previous sections, the PDF of the mass density contrast is calculated assuming a lognormal distribution (equation~\ref{eq_logn}), specified by the adopted cosmology. The values of measured biasing parameters are presented in Table~\ref{tab_biaspar}. The detailed discussion and interpretation of the results obtained from the biasing analysis is presented in the following subsections. When interpreting the zCOSMOS biasing results and comparing them to other work, one has to keep in mind that the obtained results and their exact redshift evolution is cosmology dependent. This is evident from the dependence of the growth rate on the cosmological parameters, and particularly the results are dependent on the chosen $\sigma_8(z=0)$ normalisation. SBD find that the linear biasing parameter $\hat{b}$ changes as $\sigma^{-1}$, while the nonlinearity parameter $\tilde{b}/\hat{b}$ changes only very weakly with $\sigma$, $\tilde{b}/\hat{b} \sim \sigma^{0.15}$. We show in Figure~\ref{fig_biassigmaval} the change of the shape of the mean biasing function with the $\sigma_8$ parameter. Lowering $\sigma_8$ will produce less structure at a given epoch, and it has a similar effect on the change in the shape of the mean biasing function as the increasing the mean intergalaxy separation.

\subsection{Shape of the conditional mean function $\langle \delta_g|\delta \rangle$}

\begin{figure}
\includegraphics[width=0.31\textwidth]{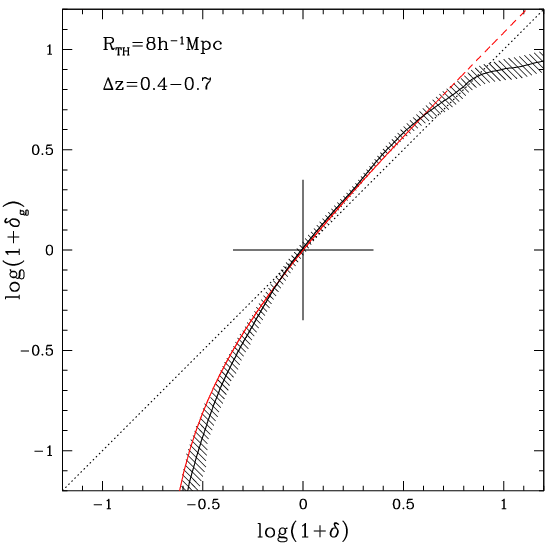}
\includegraphics[width=0.31\textwidth]{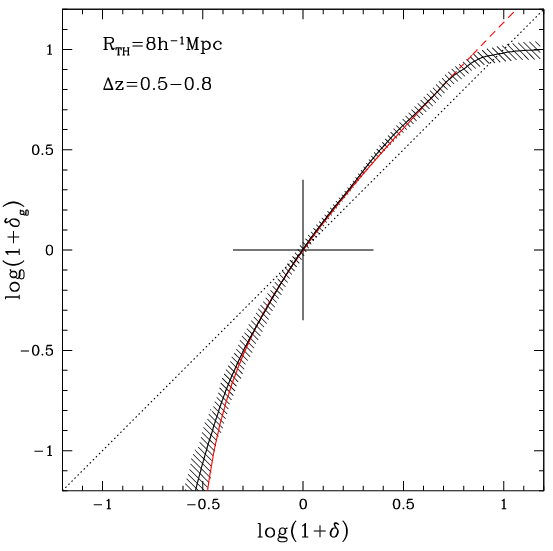}
\includegraphics[width=0.31\textwidth]{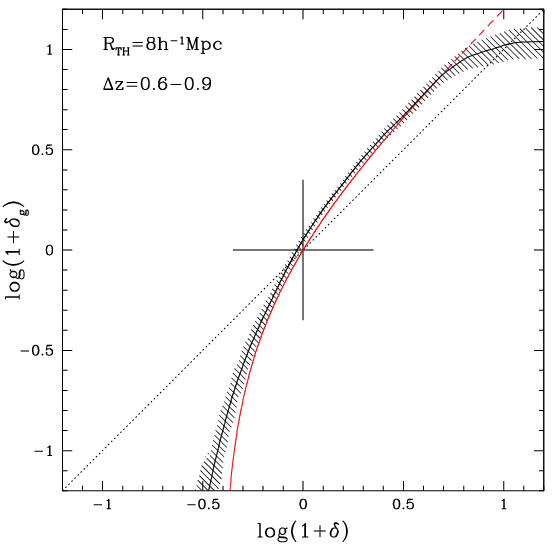}
\caption{\label{fig_bias_zth8} Conditional mean function $\langle \delta_g|\delta \rangle$ for the density field of the zCOSMOS galaxies ($M_B<-20-z$) obtained by smoothing on scales of 8 \hh Mpc. The red curve corresponds to the linear biasing case $\delta_g=\hat{b}\delta$. The different panels are for the different redshift bins: $0.4<z<0.7$, $0.5<z<0.8$ and $0.6<z<0.9$ from the left to the right, respectively.}
\end{figure}

\begin{figure}
\includegraphics[width=0.31\textwidth]{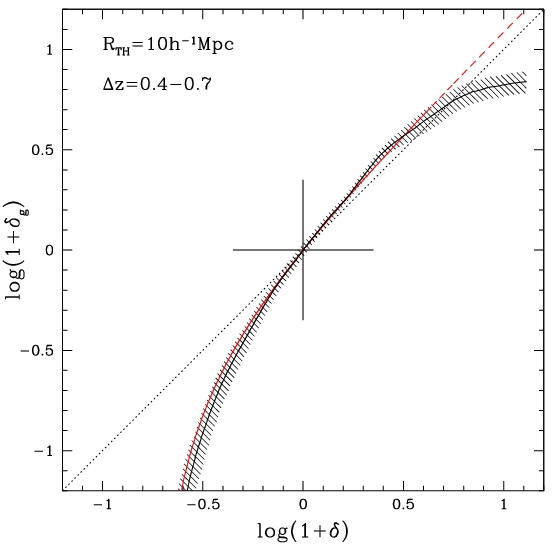}
\includegraphics[width=0.31\textwidth]{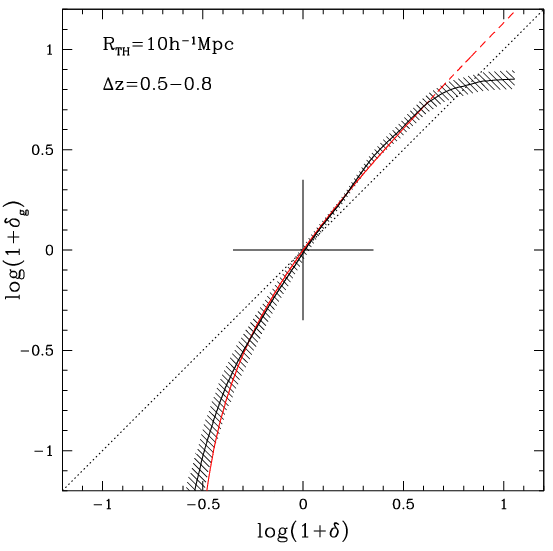}
\includegraphics[width=0.31\textwidth]{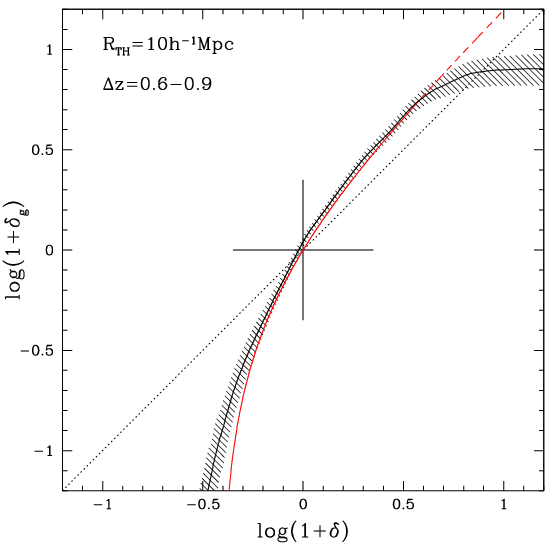}
\caption{\label{fig_bias_zth10} Conditional mean function $\langle \delta_g|\delta \rangle$ for the density field of the zCOSMOS galaxies ($M_B<-20-z$) obtained by smoothing on scales of 10 \hh Mpc. The meaning of the curves and symbols is as in Figure~\ref{fig_bias_zth8}.}
\end{figure}

We show the conditional mean function $\langle \delta_g|\delta \rangle$
 of  galaxies in Figures~\ref{fig_bias_zth8}  and~\ref{fig_bias_zth10}, where the   galaxy  overdensity fields 
are reconstructed  with the luminosity complete $M_B<-20-z$  sample of galaxies  for  the  smoothing  filters  of  8  and  10  \hh  Mpc,
respectively. We  also present
the corresponding  linear   biasing   approximation  $\delta_g=b_L   \delta$ 
 with
$b_L=\hat{b}$ at every $\delta$.

The $\langle \delta_g|\delta \rangle$ function vanishes in the most underdense regions. At moderate underdensities the $\langle \delta_g|\delta \rangle$ function sharply rises  as we approach $\delta_g  \approx \delta  =  0$. Starting in these and up to the mildly overdense regions, the $\langle \delta_g|\delta \rangle$ function closely follows the linear relation with $\delta$. In the most overdense regions, our  results
suggest that galaxies are antibiased tracers of the underlying matter
distribution.   The local  slope $b(\delta, z, R)$ of  the biasing  relation  in the  underdense regions  is
larger than unity. In the overdense regions, the trend is less clear, as the local slope 
can  take values  both  larger  and smaller  than  unity.

This characteristic shape of the conditional mean function $\langle \delta_g|\delta \rangle$ persists for all the samples of tracer galaxies and redshift intervals covered by our study. The biasing relation between galaxies and  matter is clearly nonlinear in the most underdense and overdense regions in $0.4<z<1$,  in agreement  with previous work at these redshifts
based  both   on  simulations  and   semi-analytical  modelling  \citep[SAM; SBD, ][]{Somerville.etal.2001} or observations \citep{Marinoni.etal.2005}.

Theoretical works provide some explanation for the observed shape of the conditional mean function $\langle \delta_g|\delta \rangle$. Vanishing of the function in the underdense regions can be interpreted within a scenario in  which galaxies do  not form  below some mass density threshold. However, as discussed above, one needs to be careful in this interpretation and take the possible shot noise effects into account before inferring this threshold. The antibiasing of  galaxies in the most overdense regions observed at redshifts  of unity and below  can be explained by  quenching of galaxy
formation  in these regions, as at these redshifts the densest regions  become too  hot  \citep{Blanton.etal.2000}.  The other possibility  to explain the antibiasing of galaxies
in the  most overdense  regions are different  epochs of  formation of
galaxies  in  overdense  and  underdense  regions  \citep{Yoshikawa.etal.2001}. This follows from the hierarchical scenario of galaxy formation, where for a given mass scale, there is a tendency for objects in overdense regions to form earlier than objects in the underdense regions. Therefore the young galaxies (e.g. with formation redshift since 1.7 in \citealt{Yoshikawa.etal.2001}) are expected to form in low density, which are also low temperature, environments. Also, the merging  of galaxies in high density environments
could lower the number density  of galaxies used to derive the density
field \citep{Marinoni.etal.2005}.

\subsection{Scale dependence}

\begin{figure}
\includegraphics[width=0.45\textwidth]{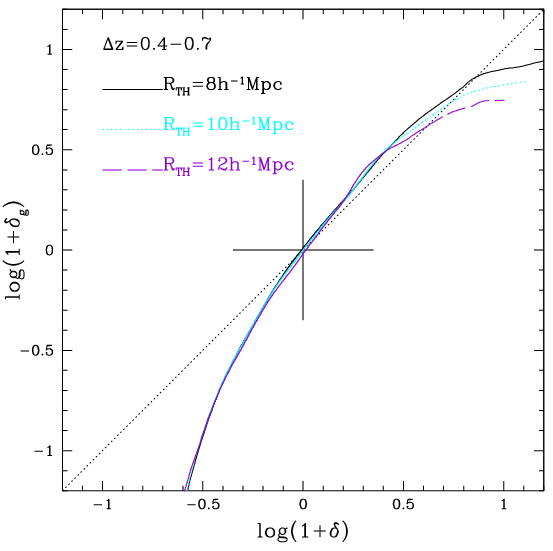}
\includegraphics[width=0.45\textwidth]{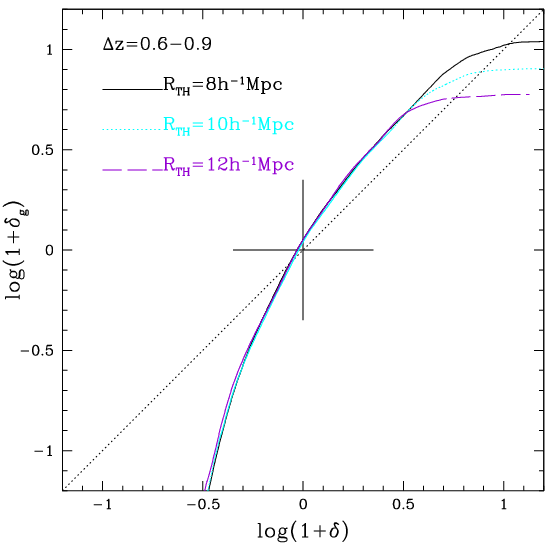}
\caption{\label{fig_biasthvar} Conditional mean function $\langle \delta_g|\delta \rangle$ for the density field of the zCOSMOS galaxies ($M_B<-20-z$) obtained by smoothing on various scales. The resulting functions are presented with the black continuous, cyan dotted and violet dashed lines for the scale of 8, 10 and 12 \hh Mpc, respectively, in two redshift intervals: $0.4<z<0.7$ (left) and $0.6<z<0.9$.}
\end{figure}

To examine the possible dependence of the shape of the conditional mean function $\langle \delta_g|\delta \rangle$ on the smoothing scale, we present in Figure~\ref{fig_biasthvar} $\langle \delta_g|\delta \rangle$ derived with the $M_B<-20-z$ tracer galaxies smoothed with three different filters: 8, 10 and 12 \hh Mpc.  The $\langle \delta_g|\delta \rangle$ functions for different smoothing scales are almost identical to each other for the largest part of the explored range of overdensities, in both redshift bins which we plot ($0.4<z<0.7$ and $0.6<z<0.9$). In the most overdense regions the characteristic $\delta$ at which galaxies become antibiased shifts from higher to lower values as the smoothing scales shift from lower to higher values.

\begin{figure}
\includegraphics[width=0.55\textwidth]{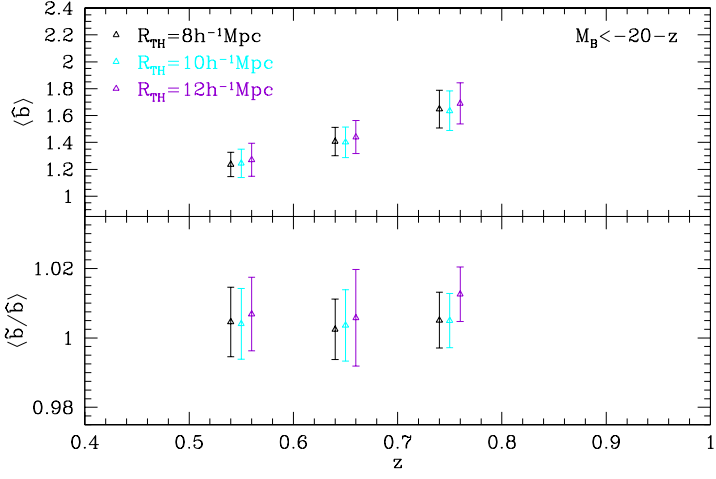}
\caption{\label{fig_biasparthvar} Biasing parameters for the 10k zCOSMOS overdensity field ($M_B<-20-z$) calculated for the various smoothing scales. The redshift evolution of the linear biasing parameter $\hat{b}$ is presented in the top panel and the redshift evolution of the nonlinearity parameter $\tilde{b}/\hat{b}$ is presented in the bottom panel. The black, cyan and violet symbols represent the parameters for the scale of 8, 10 and 12 \hh Mpc, respectively. The errors are calculated as the standard deviation of the corresponding biasing parameters from 12 40k-type mocks. Some points and their errors are displaced along redshift-axis from the mean redshift in the bin where analysis was carried out for the sake of clarity.}
\end{figure}

We summarise the impact of the different smoothing scales on the biasing parameters in Figure~\ref{fig_biasparthvar}, using the same overdensity field. We do not detect any significant dependence of  the derived linear biasing parameter nor the nonlinearity parameter on the smoothing scale. However,  we cover only a very narrow range in scales and taking the estimated errors into account, we would detect a dependence of the bias on the smoothing scale only if the effect is very strong.

The negligible dependence of the biasing function and parameters on the scales of 8 \hh Mpc and larger is in agreement with the arguments from a number of theoretical works, according to which bias is expected to be constant on scales larger than a few \hh Mpc \citep[e.g.][]{Kauffmann.etal.1997, Mann.etal.1998, Benson.etal.2000}. However, using hydrodynamical simulations \citet{Blanton.etal.1999} detect decreasing of galaxy bias with the smoothing scale, where bias is defined as the ratio between the variances of the number of galaxies and mass withing spheres of radius $R$. This dependence is significant on the scales $R$ smaller than the transition scale between the linear and nonlinear regimes (and it is 16 \hh Mpc in \citealt{Blanton.etal.1999} simulations). \citet{Blanton.etal.1999} explain the scale dependence of the bias to follow from the dependence of the galaxy density field on the local temperature, which reflects the gravitational potential related to the mass density field. If the gas is too hot, galaxies will not form, influencing directly the reconstruction of the galaxy density field.

\subsection{Luminosity dependence}

\begin{figure}
\includegraphics[width=0.45\textwidth]{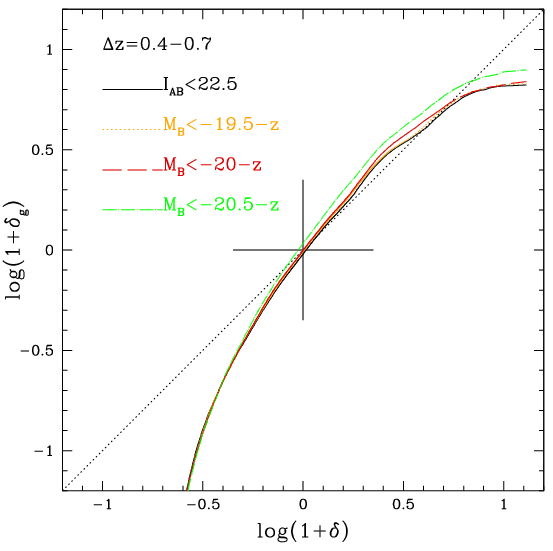}
\includegraphics[width=0.45\textwidth]{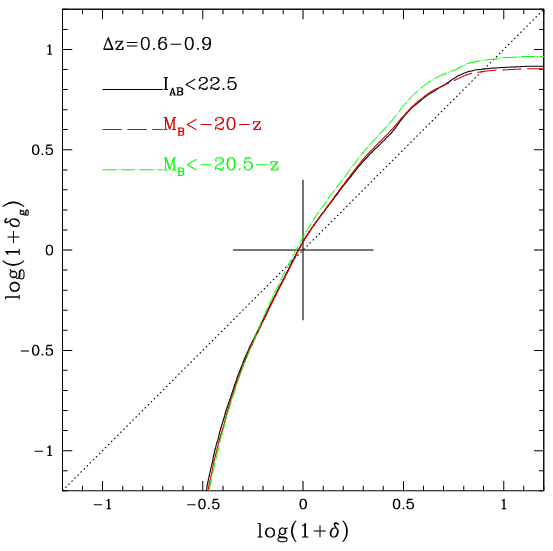}
\caption{\label{fig_biasmBvar} Conditional mean function $\langle \delta_g|\delta \rangle$ for the density field of the zCOSMOS galaxies of various luminosity thresholds obtained by smoothing on a scale of 10 \hh Mpc. The resulting functions are presented with the black continuous, yellow dotted, red dashed and green dot-dashed lines for the samples of $I_{AB}<22.5$, $M_B<-19.5-z$, $M_B<-20-z$ and $M_B-20.5-z$ galaxies, respectively, in two redshift intervals: $0.4<z<0.7$ (left) and $0.6<z<0.9$ (right). For the luminosity complete samples, only results for the samples for which we are complete in a given redshift interval are presented.}
\end{figure}

In this subsection we investigate the dependence of the $\langle \delta_g|\delta \rangle$ function and biasing parameters on the luminosity of galaxies used to reconstruct the overdensity field, for a fixed smoothing scale of 10 \hh Mpc. The $\langle \delta_g|\delta \rangle$ function derived with the tracer galaxies of different luminosity thresholds in two redshifts bins ($0.4<z<0.7$ and $0.6<z<0.9$) is presented in Figure~\ref{fig_biasmBvar}. We detect a weak dependence of $\langle \delta_g|\delta \rangle$ on the luminosity of the tracer galaxies. This effect is clearest in the overdense and mildly underdense regions of the mass density contrast distribution, where the local bias of more luminous galaxies is higher than the local bias of less luminous galaxies. In the most underdense regions the differentiation between the conditional mean functions $\langle \delta_g|\delta \rangle$ of the galaxies with different luminosity thresholds becomes barely visible; however, in this regime $\langle \delta_g|\delta \rangle$ of less luminous galaxies is systematically above $\langle \delta_g|\delta \rangle$ of more luminous galaxies, although hard to see. Considering the reconstruction errors, the differentiation between the $\langle \delta_g|\delta \rangle$ functions due to the luminosity of tracer galaxies should be even more pronounced. We have seen in Figures~\ref{fig_meanbiasrecflux} and~\ref{fig_meanbiasrecmB} that due to the reconstruction errors the conditional mean function $\langle \delta_g|\delta \rangle$ in the regions of the lowest mass density contrast is getting shifted in the opposite directions for the flux limited and luminosity complete samples, making galaxies to appear higher or lower biased for these two samples, respectively. Taking these errors into account, we expect the difference between the critical $\delta$ below which $\delta_g$ of galaxies do not trace the matter any more to be about $\log(1+\delta) \sim 0.1$ for the $\langle \delta_g|\delta \rangle$ function of the flux limited and $M_B<-20-z$ samples of galaxies.

The $\langle \delta_g|\delta \rangle$ function in \citet{Marinoni.etal.2005} shows the same behaviour with luminosity in the most overdense regions, even though their errors in the most overdense regions are larger. Interestingly, \citet{Marinoni.etal.2005} detect a much clearer differentiation of the conditional mean function $\langle \delta_g|\delta \rangle$ with luminosity in the regions of $\delta < 0$, in the same direction as our data indicate. They interpret the change in the shape of the $\langle \delta_g|\delta \rangle$ function in the underdense regions with luminosity as an indication that the formation efficiency of galaxies is shifting towards higher densities with increasing luminosity. However, \citet{Marinoni.etal.2005} do not take into account the change in the mean intergalaxy separation between different populations of galaxies, as this fact by itself will cause this type of shift in the $\langle \delta_g|\delta \rangle$ function (due to the different shot noise manifestation for the different populations of galaxies as shown in Section~\ref{sub_shotnoise}).

\begin{figure}
\includegraphics[width=0.55\textwidth]{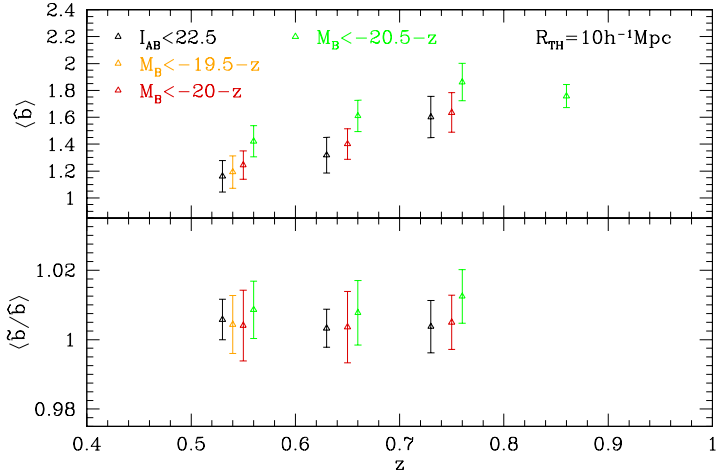}
\caption{\label{fig_biasparmBvar} Biasing parameters for the 10k zCOSMOS overdensity field of the various luminosity thresholds. The black, yellow, red and green symbols represent the parameters for the samples of $I_{AB}<22.5$, $M_B<-19.5-z$, $M_B<-20-z$ and $M_B-20.5-z$ galaxies, respectively. For the luminosity complete samples, only results for the samples for which we are complete in a given redshift interval are presented. Details are as in Figure~\ref{fig_biasparthvar}.}
\end{figure}

The mean biasing parameter measured by $\hat{b}$, presented in the upper panel of Figure~\ref{fig_biasparmBvar}, shows a systematic dependence on the luminosity of the tracer galaxies: it is higher for the more luminous galaxies. The errors in the plot include only the cosmic variance and they indicate the expected spread in the obtained $\hat{b}$ values at a given redshift. The errors from the reconstruction can be neglected at lower redshifts, but for  $z > 0.7$ they artificially increase the $\hat{b}$ parameter for both the flux limited and luminosity complete samples  (for about $\sim 5\%$ and $\sim 4\%$ at $z \sim 0.75$ for the flux limited and $M_B<-20-z$ luminosity complete samples, respectively, using a top-hat smoothing of 8 \hh Mpc, see Figure~\ref{fig_biasparrec}). In addition, more luminous samples of galaxies will have larger mean intergalaxy separation, what will artificially increase the $\hat{b}$ parameter. For example, as seen in the mocks, the relative increase in $\hat{b}$ (with respect to the results with $l \sim 2.7$ \hh Mpc) would be $\sim4.5\%$ and $\sim 10.5\%$ at $z \sim 0.75$ for $M_B<-20-z$ and $M_B<-20.5-z$ samples of galaxies, purely because the differences in the mean intergalaxy separations. Given that we measure larger differences between the $\hat{b}$ parameters from different luminosity complete samples than expected by purely taking into account mean intergalaxy separations for these different populations, we are confident that the trend of increasing $\hat{b}$ with luminosity reflects intrinsic physical processes in galaxy formation. The dependence of the correlation function on the luminosity is a known observational result (e.g. \citealt{Zehavi.etal.2005, Pollo.etal.2006, Coil.etal.2008}; but see \citealt{Meneux.etal.2009}), therefore an increase in the linear bias with luminosity is expected.

We do not detect any significant dependence of nonlinearity on the luminosity of the tracer galaxies. Based on the mocks, the errors from the density field reconstruction can be neglected  in the nonlinearity parameter. The shot noise effects in the nonlinearity parameter tend to be more important at higher redshifts. For example, the shot noise effect would be responsible for about 0.3$\%$ decrease in the nonlinearity for the  mean intergalaxy separation 7-8 \hh Mpc.

\subsection{Redshift evolution} 

The evolution of the conditional mean function $\langle \delta_g|\delta \rangle$ and biasing parameters with redshift can be studied in all results presented so far. At a given luminosity and smoothing scale, the basic shape of the $\langle \delta_g|\delta \rangle$ function (e.g. Figures~\ref{fig_bias_zth8} and~\ref{fig_bias_zth10}) is preserved in all three redshifts bins probed by the overdensity field traced by the 10k zCOSMOS galaxies. The characteristic value of $\delta$ at which galaxies (apparently) do not trace matter any more, defined here by $\delta_g \le -0.9$,  shifts from $\delta \approx -0.70$ at $z \sim 0.55$ to $\delta \approx -0.63$ at $z \sim 0.75$ for the $M_B<-20-z$ tracers. However, this difference is comparable to the reconstruction error in the $\langle \delta_g|\delta \rangle$ function with this sample (Figure~\ref{fig_meanbiasrecmB}), therefore we can not distinguish whether the redshift evolution in this characteristic $\delta$ value is the real effect or not. On the other hand, at positive $\delta$ galaxies become antibiased at gradually   higher matter overdensities when going from lower to higher redshifts. This value of $\delta$ shifts from 4.46 at $z \sim 0.55$  to 6.75 at $z \sim 0.75$. Given that we have selected partially overlapping redshift slices in order to reduce cosmic variance, we have also smoothed the possible redshift evolution in the shape of the $\langle \delta_g|\delta \rangle$ function.

The redshift evolution of the mean biasing parameter $\hat{b}$ is more evident (Figures~\ref{fig_biasparthvar} and ~\ref{fig_biasparmBvar}). For the adopted cosmology, $\hat{b}$ increases from $1.24\pm 0.11$ at $z \sim 0.55$ to $1.63\pm 0.15$ at $z \sim 0.75$ for the $M_B<-20-z$ sample of galaxies for the top-hat smoothing of 10 \hh Mpc, taking very similar values for the other smoothing scales. Also, $\hat{b}$ is higher for the more luminous samples of galaxies (but see the  discussion on errors in the previous Subsection).

The nonlinearity of the mean biasing function as measured by $\tilde{b}/\hat{b}$ is the least affected by the shot noise and reconstruction errors. The $\tilde{b}/\hat{b}$ do not show  any significant redshift evolution, in addition to no dependence on the smoothing scale or on the luminosity threshold of galaxies used to reconstruct the overdensity field. The nonlinearity parameter is of the order of a few percent ($\sim 2\%$).

The redshift evolution of the shape of the $\langle \delta_g|\delta \rangle$ function in the manner indicated by our results has been seen in both the simulations and SAMs \citep[e.g.][]{Somerville.etal.2001} and observations \citep{Marinoni.etal.2005}. The redshift evolution of the biasing is visible most clearly using the linear biasing parameter. The increase in the linear biasing parameter with redshift seen in the zCOSMOS data supports the theoretical predictions of the bias evolution. \citet{Blanton.etal.2000} discriminate between three different effects which are responsible for this bias behaviour. First, galaxy formation process shifts from the highest peaks in the density field at early epoch to the lower peaks in the density field as time progresses, i.e. this process shifts from the most biased to less biased tracers of the underlying density field. Second, the formation of galaxies in the most dense environments is halted towards lower redshifts, because these regions become filled with gas which is shock heated and virialised, but which is not able to cool and collapse. At  higher redshifts ($z \sim 3$)  galaxies are expected to
be  biased  even in  the  highest  density  regions, because they are  still
sufficiently cold enough to allow for fast cooling of  the  clumps  of  gas. These two effects combined are responsible for the shift in the galaxy formation from the most dense to less dense environments. Third, once galaxies are formed, they experience the same gravitational physics as the dark matter, and therefore the distribution of galaxies and matter becomes more and more similar \citep{Fry.1996, Tegmark&Peebles.1998}.

\subsection{Biasing for the luminosity or stellar mass weighted density field} 
\label{sub_biaswlm}

The general equation to reconstruct the galaxy density field, Equation~\ref{eq_rhodef}, is commonly used by weighting all galaxies in the tracer sample equally, i.e. using $m_i = 1$. However, it can be expected that some galaxy properties, particularly total mass, are better tracers of the underlying matter density field. To address this quantitatively, we repeat the galaxy density field reconstruction in the zCOSMOS volume using the sample of $M_B<-20-z$ and $M_B<-20.5-z$ tracer galaxies as defined in Subsection~\ref{sub_densrec}, but now using two different $m_i$-weighting schemes. We use $m_i = L_{B,i}$ and $m_i = M_{*,i}$, which are estimated reasonably well for practically all galaxies in the 40k sample (see \citealt{Oesch.etal.inprep} and  \citealt{Pozzetti.etal.inprep} for the absolute magnitude and stellar mass measurements).

\begin{figure}
\includegraphics[width=0.45\textwidth]{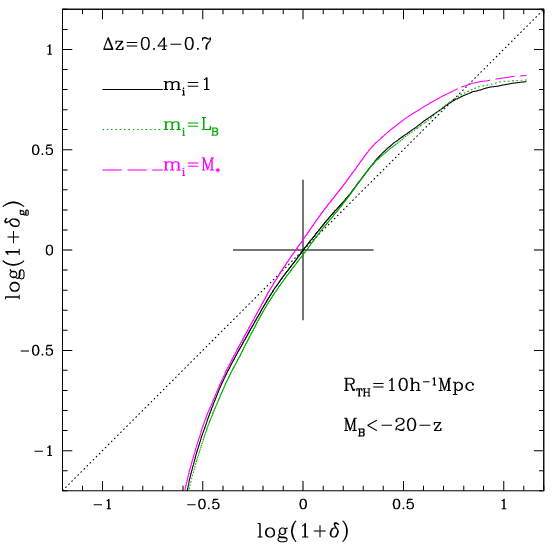}
\includegraphics[width=0.45\textwidth]{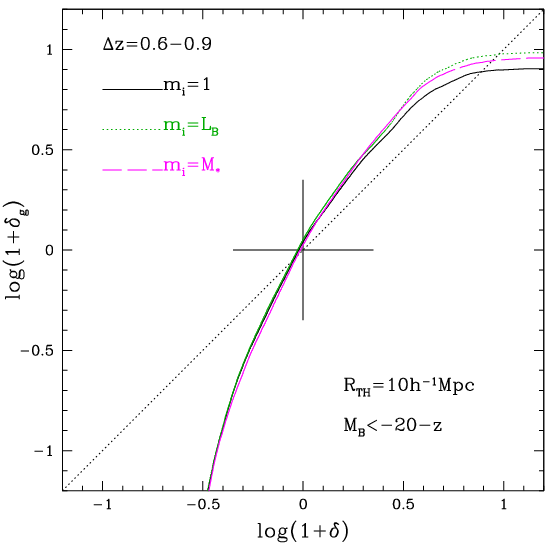}
\caption{\label{fig_biaswvar} Conditional mean function $\langle \delta_g|\delta \rangle$ for the density field of the zCOSMOS galaxies ($M_B<-20-z$) obtained by smoothing on a scale of 10 \hh Mpc with the different weighting schemes $m_i$ (see Equation~\ref{eq_rhodef}) for the density field reconstruction. The resulting functions are presented with the black continuous, green dotted and magenta dashed lines for the $m_i=1$, $m_i=L_{B,i}$ and $m_i=M_{*,i}$ weighting schemes, respectively, in two redshift intervals: $0.4<z<0.7$ (left) and $0.6<z<0.9$ (right).}
\end{figure}

The conditional mean function $\langle \delta_g|\delta \rangle$ between the $L_B$- and $M_*$-weighted galaxy overdensity field and the matter overdensity field for the $M_B<-20-z$ sample of tracer galaxies is presented in Figure~\ref{fig_biaswvar} for two redshift bins ($0.4<z<0.7$ and $0.6<z<0.9$). We have not repeated the error estimates from the  mock catalogues for a different $m_i$-weighting. The $\langle \delta_g|\delta \rangle$ functions of the differently $m_i$-weighted galaxy overdensity field are consistent within the $1\sigma$ errors of the $m_i=1$ weighted galaxy density field over the whole $\delta$-interval probed. There is some indication that at positive $\delta$, the galaxy density field weighted by $m_i=M_{*,i}$ is more biased at $z \sim 0.55$, while  the galaxy density field weighted by $m_i=L_{B,i}$ is more biased at $z \sim 0.75$ than the other considered types of galaxy density field.

\begin{figure}
\includegraphics[width=0.55\textwidth]{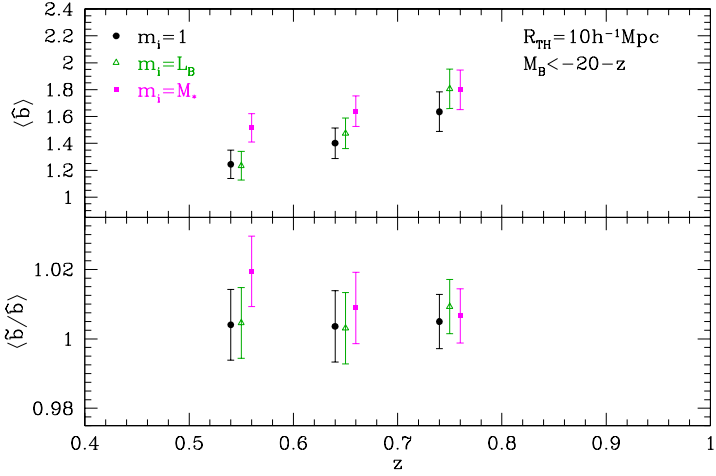}
\caption{\label{fig_biasparwvar} Biasing parameters for the 10k zCOSMOS overdensity field ($M_B<-20-z$ and $R=10$ \hh Mpc) calculated with the different weighting schemes $m_i$ (see Equation~\ref{eq_rhodef}) for the density field reconstruction. The black circles, green traingles and pink squares represent the parameters for the samples of $m_i=1$, $m_i=L_{B,i}$ and $m_i=M_{*,i}$ weighting schemes. Details are as in Figure~\ref{fig_biasparthvar}.}
\end{figure}

There seems to be a clear difference in the biasing results obtained with the different $m_i$-weights, when characterising the mean biasing function by its moments, visible in Figure~\ref{fig_biasparwvar}. In the absence of the measured errors for the various $m_i$-weightings, we have adopted the errors on the biasing parameters  estimated for the $m_i=1$ weighting. They can be considered at least as an indication for the real errors. The $\hat{b}$ parameter increases with increasing $z$ for all three types of adopted $m_i$-weighting, hinting at a different speed in this evolution. The $\hat{b}$ parameters of the $L_B$- and $M_*$-weighting are very similar and larger than $\hat{b}$ for the $m_i=1$ weighting at $z \sim 0.75$. The $\hat{b}$ parameter of the $L_B$-weighting decreases more rapidly and at $z \sim 0.55$ is consistent with $\hat{b}$ parameter of the $m_i=1$  weighting. The $\hat{b}$ parameter of the $M_*$-weighting is either larger or the same within the errors compared to the $\hat{b}$ parameter of the $L_B$- or unity-weighting. This happens due to higher $M_*/L_B$ ratio in high density regions progressively appearing as $z$ decreases (see \citealt{Bolzonella.etal.inprep} for the evolution of the zCOSMOS stellar mass function in different environments).

The nonlinearity parameter $\tilde{b}/\hat{b}$ does not evolve with redshift for $m_i=L_{B,i}$, as we have seen already for $m_i=1$, and it is consistent for these two galaxy density field reconstructions. There is some evidence that  $\tilde{b}/\hat{b}$ with $m_i=M_{*,i}$ weighting increases with decreasing $z$, reaching the value of 1.02 at $z \sim 0.55$. That is the highest value of the nonlinearity parameter for the $M_B<20-z$ sample and $R_{TH}=10$ \hh Mpc smoothing  at all redshifts probed by the zCOSMOS.

The behaviour of the $\langle \delta_g|\delta \rangle$ function and the biasing parameters with the different $m_i$-weighting schemes for $M_B<-20.5-z$ tracer galaxies in $0.4<z<0.9$ is fully consistent with the equivalent results derived with $M_B<-20-z$ tracer galaxies (described in this section).

\section{Comparison to previous work}
\label{sec_comparison}

The zCOSMOS biasing results are in qualitative agreement with the biasing framework extracted from the $\Lambda$CDM models and with a previous study of similar type based on the VVDS data. In order to put some tighter constraints on the models of galaxy formation and evolution it is important to carry out more exact comparisons with previous results. For this purpose, we use only the linear biasing parameter, which is much easier to compare.

\subsection{Biasing parameter from the nonlinear biasing analysis}

\begin{figure}
\includegraphics[width=0.45\textwidth]{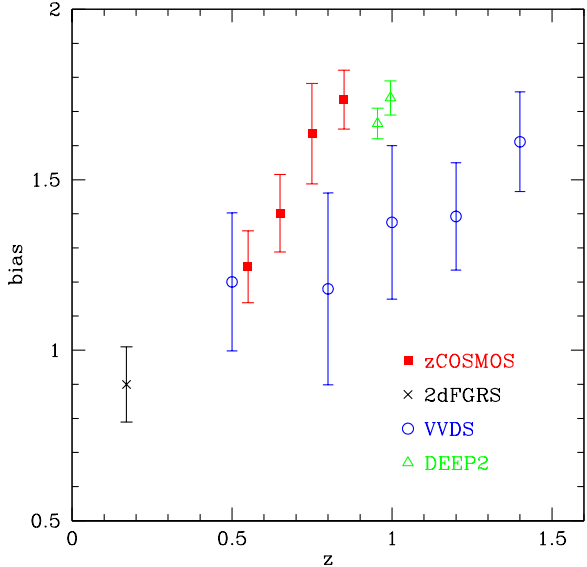}
\caption{\label{fig_biaslitrcomp}Comparison of the zCOSMOS linear biasing parameters to the bias values from similar analyses available in the literature. The filled squares correspond to the zCOSMOS measurements of $\hat{b}$, the empty circles are $\hat{b}$ values based on the nonlinear biasing analysis in the VVDS (Marinoni et al. 2005) and the cross is the $b_1$ value from the 2dFGRS (Verde et al. 2002). The zCOSMOS points are presented at the mean redshift of the bins $\Delta z=0.3$ wide. The three lower $z$ points are calculated for the $M_B<-20-z$ sample, while the $z \sim 0.85$ point is for the $M_B<-20.5-z$ sample and the smoothing scale is $R=10$ \hh Mpc for all the points. The biasing values from the VVDS are inferred from the overdensity field reconstructed using a sample with $M_B<-20.77$ on $R=5$ \hh Mpc scale in $0.4<z<0.7$, and on $R=10$ \hh Mpc scale in $0.7<z<0.9$, $0.9<z<1.1$, $1.1<z<1.3$ and $1.3<z<1.5$. The VVDS points are plotted at the centre of the  corresponding bin, with the exception of the lowest $z$ point, offset along the redshift axis due to clarity by $-0.05$. The $b_1$ value from the 2dFGRS is recalculated for $M_B<-20-z$ galaxies at $z=0.17$, which is the effective redshift of 2dFGRS. For a comparison, we add linear bias values obtained from the clustering statistics in the DEEP2 \citep{Coil.etal.2006}, represented as triangles. The DEEP2 bias is plotted at the mean redshift of the $0.75<z<1.2$ interval used for the analysis, offset along the redshift axis due to clarity by $-0.02$ for the $M_B<-20.77$ sample and by $+0.02$ for the $M_B<-21.27$ sample of galaxies. See text for more details.}   
\end{figure}

In the study of the (nonlinear) biasing of the VVDS galaxies, \citet{Marinoni.etal.2005} assumed a cosmological model described by $\Omega_{m,0}=0.3$, $\Omega_{\Lambda,0}=0.7$ and $\sigma_8=0.9$. Moreover, they use the $\tilde{b}$ parameter  as the linear biasing parameter. For a proper comparison with our linear biasing parameter $\hat{b}$, we infer $\hat{b}$ from the values of $\tilde{b}$ and nonlinearity parameters published in \citet{Marinoni.etal.2005} and correct it for the difference in the used $\sigma_8$ normalisation, following the correction given by SBD. The comparison of linear biasing parameters from the zCOSMOS and VVDS is presented in Figure~\ref{fig_biaslitrcomp}. We use the errors for the $\tilde{b}$ parameter in \citet{Marinoni.etal.2005} as a proxy for the errors of the $\hat{b}$ parameter which we have inferred, rescaled for the difference in $\sigma_8$. Even with this effort our comparison to the results from the VVDS is still only approximate. Complications arise from the different luminosity complete samples used for the analyses. Moreover, \citet{Marinoni.etal.2005} use a non-evolving magnitude to define the samples of tracer galaxies, while we use an evolving magnitude cut.

As a local reference in Figure~\ref{fig_biaslitrcomp}, we use the bias obtained from the bispectrum of the 2dFGRS galaxies measured by \citet{Verde.etal.2002}. We take $b_1$ values (following the \citealt{Fry&Gaztanaga.1993} bias description, see Equation~\ref{eq_biasfg}) as a proxy for the linear biasing parameter, derived for $L^*$ galaxies. \citet{Norberg.etal.2001} detected a clear increase of the biasing parameter $b_{\xi}$ (from the clustering analysis) with luminosity $L$ of galaxies, described well by $\frac{b_{\xi}}{b^*_{\xi}} = 0.85 + 0.15 \frac{L}{L^*}$, where $b^*_{\xi}$ is the bias for the $L^*$ galaxies. Using this relation, we calculate the bias of the sample of $M_B<-20-z$ galaxies at the effective redshift of 2dFGRS survey $z=0.17$ \citep{Verde.etal.2002}, taking for $L$ the median luminosity of all $M_B<-20-z$ zCOSMOS tracer galaxies. We use the prescription from \citet{Norberg.etal.2002} for the transformation from our $B$ to the 2dFGRS $b_j$ filter at the median $B-V$ colour of $M_B<-20-z$ zCOSMOS tracer galaxies with secure redshifts.

It is clear that the linear bias values measured from the zCOSMOS and VVDS surveys at $z > 0.4$ are higher than the linear bias measured in the local universe. At $z \sim 0.55$, the mean redshift of the lowest explored bin of both the zCOSMOS and the VVDS surveys, there is an excellent agreement between the biasing parameters from the two surveys. However, the evolution of the linear biasing seems to be happening at different speed at higher redshifts. While in zCOSMOS we detect a constant increase in the linear biasing parameter of about $0.15-0.2$ per $\delta z = 0.1$ for the galaxies of the similar evolved luminosity, the VVDS results suggest a lower increase of the linear biasing parameter of about $0.1$ or less per $\delta z = 0.1$. Considering the errors, the difference between the two results is $\lesssim 2 \sigma$ in $0.5<z<1$.

The most probable explanation of this difference, apart from the various techniques used for the density field reconstruction, cosmological density parameters, galaxy samples etc, lies in the observed fields themselves. For instance, the COSMOS volume is dominated by large density fluctuations over the full redshift range probed by the zCOSMOS. Large structures dominate the whole zCOSMOS field even at redshifts of $z \sim 0.9$ \citep{Kovac.etal.2009}. When compared to the mock catalogues, the size of structures in the zCOSMOS galaxy density field points out that the COSMOS field is in the upper tail of the cosmic variance distribution \citep{Kovac.etal.2009}, indicated also by the other studies \citep{McCracken.etal.2007, Meneux.etal.2009}. On the other hand, the VVDS field is in the lower tail of the cosmic variance distribution \citep{Meneux.etal.2009}. Moreover, as we have mentioned already, the field used in the VVDS analysis is smaller than the zCOSMOS field and the uncertainty in the redshift precision in the VVDS is about 3 times lower than in the zCOSMOS. Nevertheless, the exact bias values and their redshift evolution should be explored further.

To complete the comparison, we add in Figure~\ref{fig_biaslitrcomp} linear bias values obtained from the clustering analysis in the DEEP2 survey in $0.75<z<1.2$ with $M_B<-20.77$ and $M_B<-21.27$ samples of galaxies in the range of projected scales $r_p=1-10$ \hh Mpc \citep{Coil.etal.2006}. These bias values fall between the zCOSMOS and VVDS measurements, however, these are bias values obtained from the different statistics. We will discuss in more details the bias obtained from the clustering anlyses in the following section.

\subsection{Biasing parameter from clustering studies}

As discussed in the Introduction, clustering studies are commonly used to derive the linear biasing parameter. One needs to keep in mind that the linear bias inferred from the clustering analysis and from the biasing moments analysis are not equivalent. In some sense, $b_{\xi}$ is differential, calculated at a given distance $r_p$, while the biasing parameter from our analysis $\hat{b}$ is some weighted average over the smoothing scale $R$. Moreover, in recent clustering studies, linear bias is derived by using simultaneously correlation values at a range of $r_p$, leading to a value of the linear biasing parameter more comparable to our approach. In addition, the difference between $\hat{b}$ and $b_{\xi}$ also reflects the physical factors such as stochasticity and nonlinearity \citep[e.g.][]{Somerville.etal.2001}. For example, at $z=0$ \citet{Somerville.etal.2001} find that $b_{\xi}$ is systematically higher than $\hat{b}$ for about $10-20\%$ for the $\Lambda$CDM cosmology for the $M_B<-20.27$ sample of (mock) galaxies.

\citet{Meneux.etal.2009} study the dependence of clustering of the 10k zCOSMOS galaxies on their luminosity for the various evolving-luminosity complete samples in $0.4<z<1$. The measured dependence of the projected correlation function on the luminosity of galaxies is very weak and without any coherent redshift evolution in the amplitude or shape. From the comparison of the correlation function of $M_B<-20.27-z$ galaxies (presented in \citealt{Meneux.etal.2009}, using here $h=0.7$ for magnitudes) to the correlation function of dark matter \citep{Smith.etal.2003} in $0.5<z<1$, these galaxies  are consistent with biasing $b_{\xi}=1.6$ at small scales, while at large scales $R > 8$ \hh Mpc the clustering analysis requires $b_{\xi} \sim 2$. The high biasing value at large scales probably reflects the relatively small transverse size of the zCOSMOS field with respect to the size of the structures, as the correlation function is not a power law (see e.g. Figure 19 in \citealt{Meneux.etal.2009}). The zCOSMOS $b_{\xi}$ inferred from the sample of $M_B<-20.27-z$ galaxies in $0.5<z<1$ at $r_p \approx 10$ \hh Mpc \citep{Meneux.etal.2009} is larger for about 24$\%$ and 15$\%$ than the $\hat{b}$ measured here for the samples of $M_B<-20-z$ (averaged over $0.5<z<0.9$) and $M_B<-20.5-z$ galaxies (averaged over $0.5<z<1$) with $R=10$ \hh Mpc, respectively. For the smaller $r_p$ scales, this $b_{\xi}$ value is larger for about 5$\%$ and it is smaller for about 6$\%$ than $\hat{b}$ for the same zCOSMOS samples as above.

It is of interest to note here that \citet{Meneux.etal.2009} compare the correlation function of the zCOSMOS and VVDS galaxies \citep{Meneux.etal.2008}, finding that the bias inferred from clustering of $\log(M_*/M_{\odot}) \ge 10$ galaxies in $0.5<z<1$ is systematically higher for the zCOSMOS than VVDS galaxies. The observed difference is fully consistent with the difference between the linear biasing parameters derived from the nonlinear biasing analysis in these two surveys, as shown here (e.g. Figure~\ref{fig_biaslitrcomp}).

Using the full DEEP2 sample, \citet{Coil.etal.2006} measure the increase of $b_{\xi}$ with luminosity on both small and large scales, where the trend is stronger on smaller scales. The measured bias $b_{\xi}$ takes values from $1.42 \pm 0.04$ for the $M_B<-19.77$ sample at $0.75<z<1$ to $1.67 \pm 0.06$ for the $M_B<-20.77$  sample at $0.75<z<1.2$ or to $1.74 \pm 0.05$ for the $M_B<-21.27$ sample of galaxies at $0.75<z<1.2$ (using here $h=0.7$ for magnitudes), derived simultaneously for  the range of scales $r_p=1-10$ \hh Mpc. We have increased the bias values and their errors quoted in \citet{Coil.etal.2006} by $\sim 13\%$ in order to correct for the different $\sigma_8$ used. The values of the most luminous DEEP2 samples of \citet{Coil.etal.2006} are comparable to our $\hat{b}$ values of the two most luminous samples at $z \ge 0.6$, and they are higher than any of the $\hat{b}$ values inferred from the VVDS sample of $M_B<-20.77$ galaxies in $0.7<z<1.5$ (see Figure~\ref{fig_biaslitrcomp}). Moreover, the DEEP2 sample is complete for the red $M_B<-21.27$ galaxies only up to $z=1.05$ \citep{Coil.etal.2008}. These galaxies are more biased than the blue galaxies \citep[e.g.][]{Coil.etal.2008} and therefore the value of bias for the $M_B<-21.27$ DEEP2 galaxies can be partially lower due to this effect (our luminosity complete samples are chosen to be also colour complete).

The comparison of the zCOSMOS linear biasing measured in this work to the previous biasing analyses based both on the clustering and moments statistics, leads us to conclude that the cosmic variance is the main contributor to the different bias values and the different rate of the redshift evolution of bias of galaxies  with similar luminosities in the existing $0.4 \lesssim z \lesssim 1.5$ spectroscopic surveys.

\subsection{Bias of the dark matter haloes}

From the observed bias of the zCOSMOS galaxies we can infer a characteristic mass of dark matter haloes that host these galaxies. Under the assumption that the observed galaxies are central galaxies of a halo, the measured bias (which as we showed does not depend on the scale of 8 to 12 \hh Mpc used to reconstruct the galaxy density field) can be matched to the bias of the dark matter haloes. With this assumption (no satellite galaxies), the infered characteristic halo mass at a given bias value will be higher than the true halo mass \citep[e.g.][]{Zheng.etal.2007}. In the zCOSMOS survey, the fraction of $I_{AB}<22.5$ galaxies in groups is increasing from $\sim 15\%$ at $z \sim 0.8$ to $\sim 25\%$ at $z \sim 0.1$ \citep{Knobel.etal.2009}, and this fraction is strongly dependent on galaxy properties \citep[e.g. stellar mass, see Figure 12 in][]{Kovac.etal.inprep}. The additional uncertainty comes from the fact that we are comparing bias of galaxies above a given luminosity with a bias of haloes of a given mass. Having in mind all the approximations mentioned above, the results on the typical halo masses of the zCOSMOS galaxies should be understood only as indicative.

We use two approximations for the bias of haloes of a given mass $M_h$ at redshift $z$ given by \citet{Sheth.etal.2001} and \citet{Pillepich.etal.2008}, to also highlight theoretical uncertainties. First, we calculate for the adopted cosmology the halo bias for the range of halo masses at redshifts 0.55, 0.65, 0.75 and 0.85, which are the mean redshifts of the redshift intervals used in our analysis. We start from the simplest scenario, in which galaxies of a given type reside always in haloes of the same mass. For the sample of $M_B<-20-z$ galaxies and $R=10$ \hh Mpc, we find by minimisation of differences between the bias of galaxies and haloes at different redshifts that the best fit mass of dark haloes to host this sample is $\sim 3 \times 10^{12}$ \hh \Msol\ or $\sim 6 \times 10^{12}$ \hh \Msol, when using the \citet{Sheth.etal.2001} or \citet{Pillepich.etal.2008} bias expressions, respectively.

\begin{figure}
\includegraphics[width=0.45\textwidth]{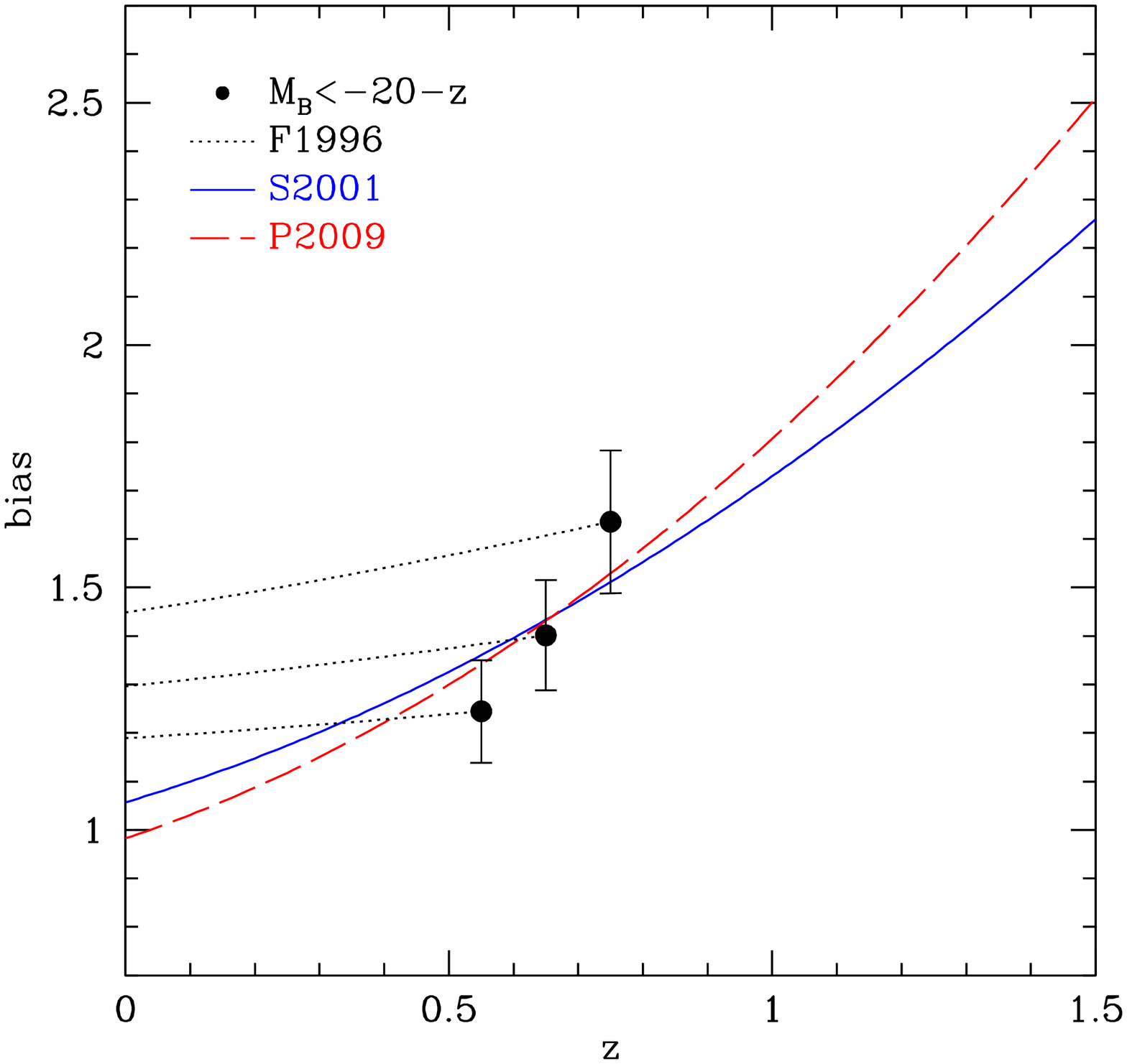}
\caption{\label{fig_biasmodelz}Redshift evolution of the bias of $M_B<-20-z$ galaxies (filled circles), $3 \times 10^{12}$ \hh \Msol\ dark matter haloes following the \citet[][S2001]{Sheth.etal.2001} expression for halo biasing (continuous line) and $6 \times 10^{12}$ \hh \Msol\ dark matter haloes following the \citet[][P2009]{Pillepich.etal.2008} expression for  halo biasing (dashed line). The evolution of biasing of haloes using the so called ``galaxy conserving''  model \citep[][F1996]{Fry.1996} is calculated assuming that at the redshift of zCOSMOS observations the model bias has the same value as the bias of $M_B<-20-z$ zCOSMOS galaxies at the $R=10$ \hh Mpc scale (dotted lines).}
\end{figure}

The redshift evolution of linear biasing of $\sim 3 \times 10^{12}$ \hh \Msol\ and $\sim 6 \times 10^{12}$ \hh \Msol\ haloes following \citet{Sheth.etal.2001} and \citet{Pillepich.etal.2008}, respectively, is presented in Figure~\ref{fig_biasmodelz}, along with the measured redshift evolution of linear bias of the $M_B<-20-z$ zCOSMOS galaxies. The models of halo biasing approximately describe the observed evolution of galaxian biasing, where the  difference falls almost completely within the $1 \sigma$ cosmic variance errors of the galaxy bias. However, our result indicates that the evolution of  galaxy bias is faster than the evolution of halo bias for the haloes of the considered mass as predicted by the both models, with \citet{Pillepich.etal.2008} bias evolution matching better the observed trend.

The considered (halo) biasing models include the effect of the merging of haloes.  For a comparison, we calculate also the evolution of biasing using the so called ``galaxy conserving''  model \citep[e.g.][]{Fry.1996}, in which the number of galaxies is preserved over cosmic time (no merging). Here, we assume that the model bias at the redshifts of zCOSMOS observations is given by the bias of $M_B<-20-z$ zCOSMOS galaxies at the $R=10$ \hh Mpc scale. The results are included in Figure~\ref{fig_biasmodelz}. The conserving model produces too high values of biasing at low $z$. This is a known result in the biasing analysis, indicating that merging is an important ingredient in the biasing of cosmic structures. Commonly, the difference in the evolution of halo and galaxy biasing (e.g. black points and red/blue curves in Figure~\ref{fig_biasmodelz}) are attributed to the different timescales of mergers of galaxies and haloes, as well as the evolution in mass-to-light ratios between haloes and galaxies \citep[e.g.][]{Somerville.etal.2001}. For the precise answer on the difference in the evolution of galaxian and halo bias, the broader baseline in redshift is needed.

\begin{figure}
\includegraphics[width=0.45\textwidth]{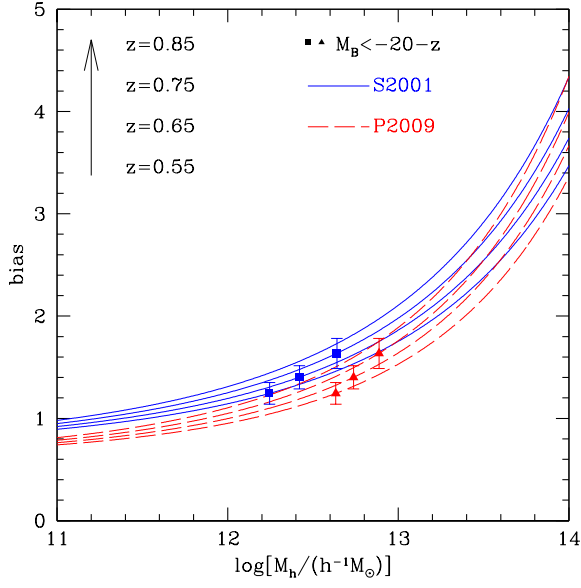}
\caption{\label{fig_biasmass}Bias of the dark matter haloes at redshifts used in this analysis. The continuous and dashed curves correspond to the bias of dark matter haloes of a given halo mass as given by \citet[][S2001]{Sheth.etal.2001} and \citet[][P2009]{Pillepich.etal.2008}, respectively. The results are calculated for redshifts $z=0.55$, $z=0.65$, $z=0.75$ and $z=0.85$ (mean redshifts of the bins used in the presented biasing analysis) for the curves presented from the bottom to the top, as indicated by the arrow along the given redshift values. The filled symbols (squares and triangles) are $\hat{b}$ values from the zCOSMOS analysis with the $M_B<-20-z$ sample on $R=10$ \hh Mpc scale, plotted at the values of a dark matter halo mass of the same bias as galaxies at the corresponding redshift.  The squares are for the \citet{Sheth.etal.2001} approximation and the triangles are for the \citet{Pillepich.etal.2008} approximation.}
\end{figure}

If we consider a different scenario,  in which the galaxies defined by their evolving $B$-band luminosity reside in haloes of different mass at different redshifts,  we find that  the characteristic mass of a halo (i.e. the halo mass at which the halo bias matches the bias of galaxies) to host a $M_B<-20-z$ galaxy increases from  $\sim 1.8 \times 10^{12}$ \hh \Msol\ at $z \sim 0.55$ to $4.3 \times 10^{12}$ \hh \Msol\ at $z \sim 0.75$ when using the halo bias approximation of \citet[][continuous curves in Figure~\ref{fig_biasmass}]{Sheth.etal.2001} or from $\sim 4.3 \times 10^{12}$ \hh \Msol\ at $z \sim 0.55$ to $7.7 \times 10^{12}$ \hh \Msol\ at $z \sim 0.75$ when using the halo bias approximation of \citet[][dashed curves in Figure~\ref{fig_biasmass}]{Pillepich.etal.2008}. This result indicates  that galaxies defined by the same $- \Delta z$ evolved luminosity reside on average in more massive haloes at higher $z$ (where $z<1$). We do not include errors in the quoted characteristic halo masses, and therefore these values should not be used to infer the exact rate in evolution in the halo mass-to-light ratio, particularly having in mind the discrepancy from the zero mean overdensity field in the $0.4<z<0.7$ bin, which can largely affect the implied evolution. Physically, the inferred trend in the halo mass-to-light ratio is consistent with a scenario in which the star formation, as traced by the $B$-band luminosity, shifts from more massive to less massive haloes with decreasing redshift. This is similar to the evolution in halo mass-to-light ratio found between the DEEP2 and SDSS \citep{Zheng.etal.2007}. Specifically, \citet{Zheng.etal.2007} find that the mean luminosity of the central galaxy increases with halo mass at both redshifts, and the central $L_*$ galaxies reside in the haloes a few times more massive at $z \sim 1$ than at $z \sim 0$. However, the physical interpretation of the results above is hampered by the fact that we (and \citealt{Zheng.etal.2007}) have used the luminosity complete samples, defined here by their evolving $B$-band luminosity, which is very sensitive to the recent star formation history. Preferentially, one should use stellar masses to define complete samples. Moreover, as seen in Section~\ref{sub_biaswlm}, the biasing inferred from the stellar mass-weighted density field is higher, and its evolution can be well described by a bias of the halo of a constant characteristic mass. The connection between galaxies and haloes inferred from the differently weighted density field has not been explored yet.

From Figure~\ref{fig_biasmass} it is also noticeable that the models for the halo bias predict higher biasing of haloes of higher mass. Moreover, the difference between the biasing of haloes of different masses increases with redshift, reflecting the fact that at high redshift more massive haloes are formed in the higher, and more rare density peaks, and therefore they are more biased tracers of the underlying mass distribution at higher redshifts. We measure higher linear biasing for more luminous galaxies, and therefore the characteristic halo mass of more  luminous galaxies is higher. For example, when using the bias expression by \citet{Sheth.etal.2001}, at $z \sim 0.55$ the characteristic mass of haloes which host $M_B<19.5-z$ and $M_B<-20.5-z$ zCOSMOS galaxies is $1.3 \times 10^{12}$ \hh \Msol\ and $3.5 \times 10^{12}$ \hh \Msol, respectively. At $z \sim 0.75$, $M_B<-20.5-z$ zCOSMOS galaxies are hosted by dark matter haloes of mass of $6.9 \times 10^{12}$ \hh \Msol. When using the bias expression by \citet{Pillepich.etal.2008}, the corresponding characterstic masses are $3.5 \times 10^{12}$ \hh \Msol, $7.1 \times 10^{12}$ \hh \Msol\ and $1.1 \times 10^{13}$ \hh \Msol\ for $M_B<19.5-z$ at $z \sim 0.55$, $M_B<-20.5-z$ at $z \sim 0.55$ and $M_B<-20.5-z$ at $z \sim 0.75$ zCOSMOS galaxies, respectively.

\section{Summary and conclusions} 
\label{sec_concl}

 In this work, we make
 use  of the reconstructed overdensity field in the zCOSMOS volume \citep[see][]{Kovac.etal.2009} to derive the conditional mean function $\langle \delta_g|\delta \rangle = b(\delta, z, R) \delta$ and the second moments of the
mean biasing function $b(\delta, z, R)$. For this purpose we employ the density field on a grid reconstructed by using the three dimensional distances between galaxies and grid points and counting the objects within a spherical top-hat aperture. We implement a novel method ZADE \citep{Kovac.etal.2009} to account for galaxies not yet observed spectroscopically in the selected samples of galaxies used to reconstruct the density field. For a biasing analysis, the main advantage of the ZADE method is that in a statistical sense, the mean intergalaxy separation is that of all galaxies in the selected galaxy sample, and not only of the sample of galaxies with spectroscopic redshifts.

We have carried out a number of tests on the mock catalogues to assess various errors which are going to affect our biasing analysis. Particularly, we have empirically estimated uncertainties due to cosmic variance, shot noise errors and the density field reconstruction errors. 
\begin{itemize}

\item Cosmic variance errors cause a spread in the $\langle \delta_g|\delta \rangle$ function: for a given $\delta$ there is a range of $\delta_g$ values measured in the mock catalogues. Quantifying this spread by the standard deviation $\sigma$ of $\log(1 + \delta_g)$, we find that $\sigma$ is largest in the most underdense regions where $\sigma \sim 0.1$, becomes lower at the intermediate $\delta$ values, $\sigma < 0.05$, and increaseas again in the most overdense regions up to $\sigma \sim 0.05$ at a given $\delta$.

\item The shot noise (discrete galaxy sampling) errors modify significantly the shape of the $\langle \delta_g|\delta \rangle$ function in the most underdense regions, making the local bias $b(\delta, z, R)$ values in the same regions to appear higher.

\item Reconstruction errors are relevant only in the underdense regions. At the reconstructed value of $\log(1+\delta_g)=-1$ in the galaxy density field, the reconstruction error can cause an uncertainty of the order of 0.1 in the matter density field $\log(1+\delta)$.  

\item The $\hat{b}$ parameter increases due to the shot noise and reconstruction errors. The $\tilde{b}/\hat{b}$ parameter is not susceptible to either of these errors. The cosmic variance causes a spread in the measured values of both of these parameters. 

\end{itemize}

We can summarise our main findings in the biasing analysis of the 10k zCOSMOS galaxies as follows:

\begin{itemize}

\item The conditional mean function $\langle \delta_g|\delta \rangle$ has a characteristic shape as described below. In most underdense regions, the mean biasing function vanishes. At some $\delta < 0$, the mean biasing function appears and then rises sharply in the underdense regions, with the local slope of the biasing function larger than unity. Starting from around mean density and towards higher overdensities, the $\langle \delta_g|\delta \rangle$ function closely follows a linear relation $\delta_g = b \delta$ with $b$ a constant. In the most overdense regions zCOSMOS galaxies are antibiased, i.e. locally $b(\delta,z,R)<1$. This is true for all samples of tracer galaxies used. The conditional mean function $\langle \delta_g|\delta \rangle$ is clearly nonlinear in the most overdense and underdense regions.

\item There is a detectable change in the shape of the $\langle \delta_g|\delta \rangle$ function with redshift in the overdense regions. For a given population,  galaxies become more biased tracers of the matter in the regions of mean and mildly positive overdensities as redshift increases from 0.4 to 1. There is an indication of an evolution in the value of $\delta$ in the overdense regions, at which galaxies become antibiased. For a given population of tracer galaxies, this happens at higher $\delta$ for higher $z$. Taking into account all the sources of errors, we cannot discriminate if the shape of the $\langle \delta_g|\delta \rangle$ function in the underdense regions stays constant or undergoes some evolution in $0.4<z<1$. It is worth stressing that redshift evolution may be not visible due to the overlapping redshift bins in the analysis.

\item The $\langle \delta_g|\delta \rangle$ function shows some dependence on the scale in the most overdense regions, more pronounced at high redshift, such that at smaller scales galaxies are more biased tracers of the underlying matter distribution. However, this is not highly statistically significant.

\item When comparing the $\langle \delta_g|\delta \rangle$ function for the different luminosity selected tracer galaxies, there is an indication that more luminous galaxies are more biased than less luminous galaxies in the regions from about mean density to the highly overdense regions.

\item The linear bias between the 10k zCOSMOS galaxies and mass is increasing with redshift. There is some evidence that more luminous galaxies are more biased tracers of matter. We do not detect any significant dependence of the linear biasing parameter on the  scale at which we measure galaxy overdensity fields. The nonlinearity parameter is of the order of a few percent. It does not change with the redshift, with the smoothing scale or with the luminosity of the tracer galaxies.

\item The linear bias of the stellar mass-weighted density field is either larger or the same within the errors compared to the linear bias of the $B$-band luminosity- or unity-weighted density field.

\item By comparing the galaxy biasing to the halo biasing, using the approximation for halo bias of \citet{Sheth.etal.2001} and \citet{Pillepich.etal.2008}, we infer that the $M_B<-20-z$ zCOSMOS galaxies in $0.4<z<1$ reside in dark matter haloes with a characteristic mass of $\sim 3$ or $\sim 6 \times  10^{12}$ \hh \Msol, using these two models, respectively. One would need to work with the stellar mass complete samples to obtain the evolution in characteristic halo mass whose physical interpretation is not ambiguos.

\end{itemize}

Broadly speaking, our results are in line with findings from the previous study of the nonlinear biasing of high redshift galaxies \citep{Marinoni.etal.2005, Marinoni.etal.2008} and, qualitatively, they follow the biasing history outlined by the theoretical works \citep[e.g.][]{Blanton.etal.1999, Blanton.etal.2000, Somerville.etal.2001}. When going into details, there are a number  of discrepancies which neeed to be solved, such as the exact ``speed'' in the evolution of the linear biasing parameter or the dependence of biasing on the luminosity of tracer galaxies. While the current biasing results from the $z<1.5$ surveys are important as they support the general picture of the biased galaxy formation and provide a framework for the future theoretical work, the fine tuning of the galaxy formation models is still hampered by the limitations of the existing  spectroscopic surveys at these redshifts, particularly by the cosmic variance.

\section{Acknowledgments}

We thank A. Pillepich for useful discussions. We thank M.G. Kitzbichler and S.D.M. White for providing the mock catalogues described in \citep{Kitzbichler&White.2007}. This work has been supported in part by a grant from 
the Swiss National Science Foundation and an ASI grant ASI/COFIS/WP3110I/ 026/07/0.


\begin{table}
\begin{tabular}{c|cc|c|cccc}
$R[h^{-1} Mpc]$ & $z_{min}$ & $z_{max}$ & $Tracers$ & $\hat{b}$ & $\Delta \hat{b}$ & $\tilde{b}/\hat{b}$ & $\Delta \tilde{b}/\hat{b}$ \\
\hline 
8 & 0.4 & 0.7 & flux & 1.15 & 0.10 & 1.005 & 0.006 \\
8 & 0.5 & 0.8 & flux & 1.32 & 0.12 & 1.002 & 0.007 \\
8 & 0.6 & 0.9 & flux & 1.61 & 0.15 & 1.004 & 0.009 \\
8 & 0.7 & 1.0 & flux & 1.63 & 0.11 & 1.003 & 0.007 \\
8 & 0.4 & 0.7 & -19.5 & 1.18 & 0.10 & 1.005 & 0.009 \\
8 & 0.4 & 0.7 & -20 & 1.24 & 0.09 & 1.005 & 0.010 \\
8 & 0.5 & 0.8 & -20 & 1.41 & 0.10 & 1.002 & 0.009 \\
8 & 0.6 & 0.9 & -20 & 1.65 & 0.14 & 1.005 & 0.008 \\
8 & 0.4 & 0.7 & -20.5 & 1.40 & 0.10 & 1.008 & 0.008 \\
8 & 0.5 & 0.8 & -20.5 & 1.58 & 0.11 & 1.007 & 0.008 \\
8 & 0.6 & 0.9 & -20.5 & 1.84 & 0.13 & 1.011 & 0.007 \\
8 & 0.7 & 1.0 & -20.5 & 1.78 & 0.08 & 1.005 & 0.006 \\
\hline
10 & 0.4 & 0.7 & flux & 1.16 & 0.12 & 1.006 & 0.006 \\
10 & 0.5 & 0.8 & flux & 1.32 & 0.13 & 1.003 & 0.005 \\
10 & 0.6 & 0.9 & flux & 1.60 & 0.15 & 1.004 & 0.008 \\
10 & 0.7 & 1.0 & flux & 1.59 & 0.11 & 1.004 & 0.006 \\
10 & 0.4 & 0.7 & -19.5 & 1.19 & 0.12 & 1.004 & 0.008 \\
10 & 0.4 & 0.7 & -20 & 1.24 & 0.11 & 1.004 & 0.010 \\
10 & 0.5 & 0.8 & -20 & 1.40 & 0.11 & 1.004 & 0.010 \\
10 & 0.6 & 0.9 & -20 & 1.64 & 0.15 & 1.005 & 0.008 \\
10 & 0.4 & 0.7 & -20.5 & 1.40 & 0.11 & 1.009 & 0.008 \\
10 & 0.5 & 0.8 & -20.5 & 1.57 & 0.12 & 1.008 & 0.009 \\
10 & 0.6 & 0.9 & -20.5 & 1.81 & 0.14 & 1.012 & 0.008 \\
10 & 0.7 & 1.0 & -20.5 & 1.73 & 0.09 & 1.007 & 0.005 \\
\hline
12 & 0.4 & 0.7 & flux & 1.19 & 0.13 & 1.010 & 0.005 \\
12 & 0.5 & 0.8 & flux & 1.36 & 0.14 & 1.005 & 0.004 \\
12 & 0.6 & 0.9 & flux & 1.66 & 0.16 & 1.011 & 0.007 \\
12 & 0.7 & 1.0 & flux & 1.62 & 0.12 & 1.008 & 0.006 \\
12 & 0.4 & 0.7 & -19.5 & 1.22 & 0.14 & 1.007 & 0.007 \\
12 & 0.4 & 0.7 & -20 & 1.27 & 0.12 & 1.007 & 0.011 \\
12 & 0.5 & 0.8 & -20 & 1.44 & 0.12 & 1.006 & 0.014 \\
12 & 0.6 & 0.9 & -20 & 1.69 & 0.15 & 1.013 & 0.008 \\
12 & 0.4 & 0.7 & -20.5 & 1.42 & 0.13 & 1.011 & 0.009 \\
12 & 0.5 & 0.8 & -20.5 & 1.61 & 0.13 & 1.012 & 0.012 \\
12 & 0.6 & 0.9 & -20.5 & 1.86 & 0.14 & 1.024 & 0.009 \\
12 & 0.7 & 1.0 & -20.5 & 1.76 & 0.09 & 1.013 & 0.005 \\
\end{tabular}
\caption{\label{tab_biaspar} Summary of the measured biasing parameters.}
\end{table}

\clearpage

\clearpage

\clearpage


\clearpage

\clearpage

\clearpage


\clearpage

\clearpage

\clearpage

\end{document}